\documentclass{iopjournal}
\usepackage{geometry}

\usepackage{amsmath} 
\usepackage{graphicx}

\usepackage[numbers,sort&compress]{natbib} 

\usepackage{ragged2e}  
\justifying  
\usepackage[english]{babel}  
\renewenvironment{abstract}
  {\begin{center}\bfseries Abstract\end{center}\justifying}
  {}
\usepackage{hyperref}
\usepackage{etoolbox} 
\usepackage{orcidlink}
\renewcommand{\footrulewidth}{0pt}  
\usepackage{caption}
\usepackage{graphicx}%
\usepackage{multirow}%
\usepackage{amsmath,amssymb,amsfonts}%

\usepackage{silence}
\usepackage{lmodern}

\usepackage[title]{appendix}%
\usepackage{xcolor}%
\usepackage{textcomp}%
\usepackage{manyfoot}%
\usepackage{booktabs}%
\usepackage{algorithm}%
\usepackage{algorithmicx}%
\usepackage{algpseudocode}%
\usepackage{listings}%
\usepackage{color}

\begin{document}


\title{Intermittent continuous-time random walks under renewal reset mechanism}

\author{ 
    Guohua Li~\orcidlink{0000-0002-2919-6807}$^{*}$,
    Hong Zhang~\orcidlink{0000-0002-7990-8980},
    Yuexiong Liu~\orcidlink{0009-0000-9893-2647}
}

\affil{School of Mathematical Sciences, Chengdu University of Technology, \\
       Chengdu 610059, Sichuan, China}

\affil{$^*$Corresponding author: \email{liguohua13@cdut.edu.cn}}

\keywords{Continuous-time random walks;  Non-Markovian resetting; Non-equilibrium stationary state; First-arrival time}

\begin{abstract}

Stochastic resetting as a practical and efficient search strategy in complex and disordered environments has long been a topic of interest to researchers.
Based on the competition between jumping and resetting, this article proposes and investigates intermittent continuous-time random walks (CTRWs) under stochastic resetting, using the smaller waiting time for jump and reset as the renewal time, where the waiting times for both jump and reset can have arbitrary distributions. After each renewal event, the system will proceed with new waiting times for jump and reset regardless of their previous histories.
We study the governing equation and Montroll-Weiss equation with renewal resetting, as well as the Markovian resetting for intermittent CTRWs. We prove the existence of non-equilibrium stationary states within the renewal reset mechanism when the jump and reset waiting times follow any exponential and power law distributions. For exponential and Gaussian distributed jump lengths, we examine the mean square displacements (MSDs) of particles to determine their monotonicity and asymptotic stability.
Moreover, we calculate the first-arrival time to quantify search efficiency, and validate the intermittent CTRWs under renewal resetting lead to a finite mean first-arrival time (MFAT) to any fixed position for exponential jump and reset waiting time distributions (WTDs), power-law jump and exponential reset WTDs, as well as exponential jump and power-law reset WTDs. However, the MFAT diverges for power-law jump and reset WTDs. The intermittent CTRW model, which is based on the competition mechanism, can be applied to many physical scenarios, such as the foraging strategy of animals that return to their nests after an unsuccessful foraging attempt, or the work planning of intelligent robots that return to energy replenishment points after prolonged operation.
\end{abstract}

\section{Introduction}
\label{s:1}

The search process is a prevalent phenomenon in nature, spanning from the microscopic scale—such as proteins seeking binding sites on DNA \cite{1ref64,ref65,ref66}—to the macroscopic scale, like animals foraging for food \cite{ref67,ref68}. In complex and disordered environments, identifying optimal search strategies to achieve efficient searching remains a core concern for researchers \cite{R2016,  PKR2020, PSS2024}. To address this challenge, stochastic resetting, defined as a search mechanism that interrupts the agent's current state and forces a return to the initial state \cite{KMSS2014, KG2015, CM2015, JPS2023, SKM2022}, has generated significant research interest.

Evans and Majumdar \cite{EM2011} pioneered the groundbreaking revelation in their paper that stochastic resetting can significantly reduce the average first-passage time of searchers and drive the system into a non-equilibrium stationary state. Subsequently, scholars have carried out an array of theoretical investigations into stochastic resetting \cite{MM2019, SGS2022, S2017, DCSM2021, MCM2019, RL2023,MC2016, ZXD2020, BP2021, PR2017, PKE2016, P2015, KP2023, PP2019, PP20190, PP2025}. Notably, leveraging the mechanism of stochastic resetting, studies have been conducted on underlying processes including anomalous diffusion \cite{MK2000} (such as CTRWs \cite{MV2013, MMSC2021}, Lévy walks \cite{ZXD2020}, and Lévy flights \cite{CM2015}) as well as scaled Brownian motion \cite{BCS2019}. The CTRW model \cite {mont65rwl2, mont69rwl3, mont73rwl4, lax73, shle74ctrw-asymp-sol, mont75, shle77pnas-ctrw-mast-eq, weis82chap-5, klaf87hdp-ctrw-stoch, weis94book, metz11, metz14, maso17ctrw, H1995, KS2011, GMV2007, SK2007, L2026} is capable of macroscopically describing complex dynamic behaviors, such as subdiffusion and localization in one-dimensional trap models \cite {BB2003}, anomalous transport in porous and heterogeneous media \cite {MR2020, AHM2019, LZZ2019, MB2000, berk02rev, SBMB2003, berk06rev-ctrw, AB2013, dorr22mob-imob}, transport in supercooled liquids \cite {HZF2014}, and epidemic spreading in heterogeneous networks \cite {ZL2020}, among others. Furthermore, in recent years, research on stochastic resetting has been extended to more complex systems and environments \cite{G2025, GB2026, GC2021}. Ref. \cite{ref69} explores how stochastic resetting aids active particles in traversing narrow pores within confined geometries; Ref. \cite{ref70} analyzes the induction of superdiffusive behavior via long-tailed resetting time distributions; and Ref. \cite{ref71} investigates the application of random resetting to explain and simulate the directed motion of Brownian particles in asymmetric periodic potential fields.

Existing research on resetting encompasses a diverse array of resetting scenarios, with various reset mechanisms exhibiting distinct characteristics due to their unique features during the reset process. First, classification can be made based on the reset time (i.e., the time interval between two consecutive reset events) within the system. If the reset time follows an exponential distribution, this form of resetting is termed Poisson resetting (or Markovian resetting)\cite{EM2011, MSM2023, MMSC2021,ref69,ref71, EMS2020}; if the reset time adheres to a general distribution (e.g., (non-Markovian) power-law distribution, gamma distribution), it is referred to as non-Poisson resetting \cite{ref70,NG2016,BS2020,SS2022}. In the latter case, the occurrence rate of reset events varies over time \cite{BEM2017, EM2016}. Second, reset mechanisms can be categorized by whether the reset is complete \cite{KG2019,BS2020}. A reset is defined as a hard reset if the position of the searcher in the system is reset and the searcher’s waiting time is restarted simultaneously; conversely, it is called a soft reset if only the searcher’s position is altered while its current waiting time state is preserved during the reset. Furthermore, based on the selection of the searcher’s reset position during the reset process, there exist reset mechanisms with historical memory, such as preferential visit resetting \cite{ref72,BEM2017} and maximum-position resetting \cite{ref74}. However, in the aforementioned body of stochastic resetting theories, resetting is often treated as an external perturbation to the system. The reset process is typically modeled as an independent, externally imposed mechanism superimposed on the underlying diffusion dynamics, with few studies constructing the system or describing the walker’s underlying dynamical evolution from the perspective of the system’s internal microscopic competition mechanism. 

Therefore, based on the research of intermittent CTRWs, this paper proposes and explores a new intermittent CTRW resetting model with a competition mechanism. Different from traditional resetting models, we regard each movement of the particle as a random competition between two potential waiting events: jumping and resetting. Specifically, each jump and reset of the particle is considered as one renewal of the system, and the waiting times for both jump and reset can have arbitrary distributions, where the renewal event of the system depends on the smaller waiting time between the two. After the renewal time, the intermittent CTRW will continue with new waiting times for jump and reset, regardless of their previous histories. In the real world, there are some physical scenarios that are compatible with this intermittent CTRW resetting model based on the competition mechanism. For example, in the foraging strategies of animals, animals must make an assessment before each movement \cite{ref75,ref76,ref77}. If an animal cannot find food in a certain area or move to the next foraging point within a certain period of time, it will determine that the current foraging path is inefficient and return to the nest to re-plan. In engineering systems, robots will evaluate their working time and energy usage time each time they perform a task \cite{ref78,ref79}. If the energy usage time is less than the required working time, the robot will return to the supply point to replenish energy before continuing to work.

In the search scenarios described above, stochastic resetting no longer represents an independent external perturbation to the diffusion system. The complexity of this search strategy stems from the diverse objective factors in real-world search scenarios, such as time, energy, and financial costs. In fact, directly modeling these search processes by quantifying these diverse objective factors remains a challenging mathematical problem, and in view of the diversity of real-world scenarios, it may lack applicability and effectiveness even when accounting for nonhomogenous (either in time or space) reset and/or jump times. To prevent searchers from operating over extended periods without regard to cost—as in the power-law distributed reset process studied by researchers (in which the mean of reset times diverges and acts as an independent external perturbation) \cite{NG2016}, leading to low search efficiency (diverging MFAT)—it is sensible and practical to implement a search strategy with a competition mechanism between jumping and resetting within the intermittent CTRW framework. The subsequent analyses in this article confirm that the intermittent CTRW model with renewal competition mechanism improves search efficiency, the exponential jump and power-law reset WTDs exhibit a finite MFAT. It demonstrates good adaptability to various search scenarios, as the specific jump and reset WTDs (regardless of the distribution type) can be derived from statistical data without the need to explicitly quantify the search costs.

In this article the intermittent CTRW model with renewal resetting is examined using the smaller waiting times for jump and reset as renewal time. The generalized Montroll-Weiss equation and the governing equation are derived for arbitrary jump length distributions and arbitrary WTDs of jump and reset. We give the probability density functions (PDFs) in Fourier space for the specified jump and reset WTDs. For the exponential distributed jump length, we derive exact analytical expressions for the stationary distribution, MSD, and MFAT (in Laplace space), thereby providing significant theoretical value for the study of renewal reset processes. For the Gaussian distributed jump length, the approximate analytical results presented here for the stationary distribution, MSD, and MFAT are of considerable practical importance.

The rest of the paper is organized as follows. We start with introducing the model of intermittent CTRW under renewal resetting using the smaller waiting times for jump and reset as the renewal time in Sec. \ref {sec-ctrw-reset}. We establish the governing equation in Sec. \ref{governing eq} and study the Markovian
resetting in Sec. \ref{Markovian reset}. we also investigate the PDFs and the stationary state in Sec.s \ref{PDF-Fourier space} and \ref{Stationary state}. Moreover, we show the main results of the typical governing equations with renewal resetting, the stationary distribution, MSD, and MFAT for the exponential and Gaussian distributed jump lengths in Sec.s \ref{Exponential distributed} and  \ref{Gaussian distributed}.
Finally, We analyze the differences in various statistical characteristics exhibited by intermittent CTRWs under different conditions and investigate the underlying causes in Sec. \ref {Comparison}. The discussion of the main theoretical results and the final conclusions are provided in Sec. \ref{sec-disc}. 

\section{Model and main equations} \label{sec-model-results}

\subsection{  Intermittent CTRWs under renewal reset mechanism} \label{sec-ctrw-reset}
In the CTRW a random walker jumps instantaneously from one site to another following a waiting period. \cite{KS2011, MK2000, MKS1998, MBK1999, SMC2018, BB2005, HC2018, C1997, CMC1997, MBK19991, MFH2010} The waiting time and the jump length as two independent identically distributed random variables  in each renewal have their joint PDF can be decoupled in simple case and coupled in complex environments for modeling anomalous transport phenomena.
The brief overview of the classical CTRW  can be found in the Appendix.

We now consider that the event for the particles jump  or reset depends on the smaller waiting time $\tau=\min\{\tau_d, \tau_s \}$. Here, $\tau_d$ and $\tau_s$
 represent independent jump and reset waiting times, whose PDFs are respectively $\psi_d(t)$ and $\psi_s(t)$. For example, if $\tau_d <\tau_s $, then it waits $\tau_d$ at $x$ and then jumps to another position.
One can see that the PDF for the event $\tau=\tau_d$ occurring can be written as:
\begin{flalign}
\begin{split}\varphi_d(t)=\psi_d(t)\Psi_s(t)=\psi_d(t)[1-\int_{0}^{t}\psi_s(t')dt'],
\label{jump-WTD-eq0}
\end{split}
\end{flalign}
where $\int_{0}^{t}\psi_s(t')dt'$ denotes the probability of reset waiting time being in $[0,t]$ and $\Psi_s(t)$ means the probability of no reset happened on the contrary.
One also see that the PDF for the event $\tau=\tau_s$ occurring can be written as:
\begin{flalign}
\begin{split}\varphi_s(t)=\psi_s(t)\Psi_d(t)=\psi_s(t)[1-\int_{0}^{t}\psi_d(t')dt'],
\label{reset-WTD-eq0}
\end{split}
\end{flalign}
here $\int_{0}^{t}\psi_d(t')dt'$ is the probability of jump waiting time being in $[0,t]$.
Besides, we find that the survival probability of the particles not moving away in $[0,t]$, neither jump nor reset as
\begin{flalign}
\begin{split}
\Psi(t)=\Psi_d(t)\Psi_s(t)=[1-\int_{0}^{t}\psi_d(t')dt'][1-\int_{0}^{t}\psi_s(t')dt'].
\label{survival-eq0}
\end{split}
\end{flalign}
In Laplace space we get with Eqs. (\ref{jump-WTD-eq0}) and (\ref{reset-WTD-eq0})
\begin{flalign}
\begin{split}
\Psi(u)=\frac{1}{u}(1-\varphi_d(u)-\varphi_s(u)).
\label{survival-eq3}
\end{split}
\end{flalign}
Here $\varphi_d(u)$ and $\varphi_s(u)$ are the Laplace transforms $t\to{u}$ of $\varphi_d(t)$ and $\varphi_s(t)$.

One then gets the renewal probability of systems depended on $\tau$ in [0, t],
\begin{flalign}
\begin{split}
P_{\tau}(t)=1-[1-\int_{0}^{t}\psi_d(t')dt'][1-\int_{0}^{t}\psi_s(t')dt'],
\end{split}
\end{flalign}
and the PDF of renewal time
\begin{flalign}
\begin{split}
\psi_{\tau}(t)&=\psi_d(t)[1-\int_{0}^{t}\psi_s(t')dt']+\psi_s(t)[1-\int_{0}^{t}\psi_d(t')dt']\\
&=\varphi_{d}(t)+\varphi_{s}(t).
\end{split}
\end{flalign}

 Firstly we consider the general case where the renewal waiting time intervals $T_n$ and jump lengths or resets (lengths) depending on position $x$ as $\xi_n$ have the joint PDF $\phi(\xi,t| x)$. We assume $\phi(\xi,t| x)$ can be written as following
\begin{flalign}
\begin{split}
\phi(\xi,t| x)=\psi_{\tau}(t)\mathbf{\Lambda}(\xi|x, \tau),
\end{split}
\end{flalign}
with the conditional distribution
\begin{flalign}
\begin{split}
\mathbf{\Lambda}(\xi|x, \tau)=\Lambda(\xi)\delta(\tau-\tau_d)+\delta(\xi+x)\delta(\tau-\tau_s),
\end{split}
\end{flalign}
that is

\begin{flalign}
\begin{split}
\phi(\xi,t| x)=\Lambda(\xi)\varphi_d(t)+\delta(\xi+x)\varphi_s(t).
\end{split}
\end{flalign}
Here the PDF $ \phi (\xi, t| x) $ is normalized, it means that when the system is reset, the particle at position $x$ goes to the 0-point position, and $\Lambda(\xi)$ is the PDF of jump lengths for diffusion.

\indent Let $I(x,t)$ be the density of particles that reach the position $x$ exactly
at time $t$. Then the evolution equation for the density of particles can be written as
\begin{flalign}
\begin{split}
\rho(x,t)=\rho_0(x)\Psi(t)+\int_0^t I(x,t-\tau)\Psi(\tau)d\tau,
\label{balance-eq20}
\end{split}
\end{flalign}
where $\rho_0(x)$ is the initial density of particles. The first term on the right-hand side of  Eq.(\ref{balance-eq20}) represents the density of particles staying at point $x$ if no jumps or resets occurred before time $t$. The second term considers the density of particles that jump to point $x$ at time $t-\tau$, provided that no new jumps or resets occur in the remaining time $\tau$. 
\noindent Taking the Laplace transform $t\rightarrow u$ and Fourier transform $x \rightarrow k$ of Eq.(\ref{balance-eq20}) yields:
\begin{flalign}
\begin{split}
\rho(k,u)=\Psi(u)\rho_0(k)+\Psi(u)I(k,u).
\label{balance-eq2}
\end{split}
\end{flalign}
\indent On the other hand, the density of particles arriving at the position $x (\neq 0)$ exactly at time $t$
obeys the diffusion balance equation:
\begin{flalign}
\begin{split}
I(x,t)=\int_{-\infty}^{\infty}\rho_0(x-\xi)\Lambda(\xi)\varphi_d(t)d\xi
+\int_0^t\int_{-\infty}^{\infty} I(x-\xi,t-\tau)\Lambda(\xi)\varphi_d(\tau)d\xi d\tau,
\label{balance-eq21}
\end{split}
\end{flalign}
The first term on the right-hand side of Eq.~(\ref{balance-eq21}) is the contribution of the particles located at position $x-\xi$ at time $t=0$ who now make their first jump to position $x$. The second term corresponds to the density of particles who arrive exactly at position $x-\xi$ at time $t-\tau$ and jump from the old position $x-\xi$ to the new position $x$ after a waiting time $\tau$. And for $x = 0$ it obeys the following diffusion and reset balance equation:  
\begin{flalign}
\begin{split}
I(0,t)=&\int_{-\infty}^{\infty}\rho_0(-\xi)\Lambda(\xi)\varphi_d(t)d\xi
+\int_0^t\int_{-\infty}^{\infty} I(-\xi,t-\tau)\Lambda(\xi)\varphi_d(\tau)d\xi d\tau\\
&+\int_{-\infty}^{\infty}\int_{-\infty}^{\infty}\rho_0(x)\delta(\xi+x)\varphi_s(t)d\xi dx
+\int_0^t d\tau\int_{-\infty}^{\infty}\int_{-\infty}^{\infty} I(x,t-\tau)\delta(\xi+x)\varphi_s(\tau)d\xi dx,
\label{balance-eq4}
\end{split}
\end{flalign}
The first two terms on the right-hand side of Eq.~(\ref{balance-eq4}) is due to the diffusion of particles to the 0-point position. 
The third term refers to the particles at all positions at the initial time that are reset to the 0-point position without diffusion.  The fourth term is the density of particles who arrive exactly at positions $x$ at time $t-\tau$ and reset from all positions $x$ to the 0-point position after a renewal waiting time $\tau$ without diffusion.
The Eq. (\ref{balance-eq4}) can be written as
\begin{flalign}
\begin{split}
I(0,t)=&\int_{-\infty}^{\infty}\rho_0(-\xi)\Lambda(\xi)\varphi_d(t)d\xi
+\int_0^t\int_{-\infty}^{\infty} I(-\xi,t-\tau)\Lambda(\xi)\varphi_d(\tau)d\xi d\tau\\
&+\varphi_s(t) + \int_0^t\left( \int_{-\infty}^{\infty} I(x,t-\tau)dx\right) \varphi_s(\tau) d\tau.
\label{balance-eq5}
\end{split}
\end{flalign}

Integrating over the variable $x$ from Eq. (\ref{balance-eq20}) we get
\begin{flalign}
\begin{split}
\int_{-\infty}^{\infty}\rho(x,t)dx=\Psi(t)\int_{-\infty}^{\infty}\rho_0(x)dx+\int_0^t \left(\int_{-\infty}^{\infty}I(x,t-\tau)dx \right)\Psi(\tau) d\tau,
\label{balance-eq6}
\end{split}
\end{flalign}
according to the normalization of probability density functions one then finds 
\begin{flalign}
\begin{split}
\Psi(t)&+\int_0^t \left(\int_{-\infty}^{\infty}I(x,t-\tau)dx\right)\Psi(\tau) d\tau=1.
\label{balance-eq7}
\end{split}
\end{flalign}
In Laplace space the above equation becomes
\begin{flalign}
\begin{split}
\int_{-\infty}^{\infty}I(x,u)dx=\frac{1}{u\Psi(u)}-1.
\label{balance-eq8}
\end{split}
\end{flalign}

\indent Making use of Eq. (\ref{balance-eq8}) from Eqs. (\ref{balance-eq21}) and (\ref{balance-eq5}) we can get the Laplace-Fourier transform $t\to{u}$ and $x\rightarrow k$ of $I(x,t)$ given by
\begin{flalign}
\begin{split}
I(k,u)=&\rho_0(k)\Lambda(k)\varphi_d(u)
+I(k,u)\Lambda(k)\varphi_d(u)+\frac{1}{u\Psi(u)}\varphi_s(u),
\label{balance-eq9}
\end{split}
\end{flalign}
that is, 
\begin{flalign}
\begin{split}
I(k,u)=\frac{\rho_0(k)\Lambda(k)\varphi_d(u)
+\frac{1}{u\Psi(u)}\varphi_s(u)}{1-\Lambda(k)\varphi_d(u)}.
\label{balance-eq10}
\end{split}
\end{flalign}

\indent Combining with Eqs. (\ref{balance-eq2}) and (\ref{balance-eq10}), after some algebraic operations we can obtain the generalized Montroll-Weiss equation for  intermittent CTRWs under renewal reset mechanism
\begin{flalign}
\begin{split}
\rho(k,u)=\frac{\rho_0(k)\Psi(u)+\frac{1}{u}\varphi_s(u)}{1-\Lambda(k)\varphi_d(u)},
\label{ML-eq0}
\end{split}
\end{flalign}
where $\rho(k,u)$, $\Lambda(k)$ and $\rho_0(k)$ are the Laplace transform $t\to{u}$ and Fourier transform $x\rightarrow k$ of $\rho(x,t)$, $\Lambda(x)$ and $\rho_0(x)$ respectively.
One then can easily check the normalization of PDF $\rho (x,t)$ via the result of $\rho(k,u)|_{k=0}=\frac{1}{u}$ in Fourier- Laplace space with Eq. (\ref{ML-eq0}).

In special case one of without reset $\psi_s(t)=0$, the Eq. (\ref{ML-eq0}) reduces to uncoupled
Montroll-Weiss equation for CTRWs
\begin{flalign}
\begin{split}
\rho_0(k,u)=\frac{\rho_0(k)\Psi_d(u)}{1-\Lambda(k)\psi_d(u)}.
\label{ML-eq01}
\end{split}
\end{flalign}

And in special case two of without diffusion $\psi_d(t)=0$, the Eq. (\ref{ML-eq0}) becomes a simple reset equation in Fourier- Laplace space
\begin{flalign}
\begin{split}
\rho(k,u)=\rho_0(k)\frac{1}{u}+(1-\rho_0(k))\frac{1}{u}\psi_s(u),
\label{}
\end{split}
\end{flalign}
with inverse Fourier-Laplace transforms that is 
\begin{flalign}
\begin{split}
\rho(x,t)=\rho_0(x)+(\delta(x)-\rho_0(x))\int_{0}^{t}\psi_s(t')dt'.
\label{}
\end{split}
\end{flalign}

\subsection{The governing equation}
\label{governing eq}
The Eq. (\ref{ML-eq0}) is rewritten as
\begin{flalign}
\begin{split}
\rho(k,u)=\rho_0(k)\Psi(u)+\frac{1}{u}\varphi_s(u) +\rho(k,u)\Lambda(k)\varphi_d(u),
\label{}
\end{split}
\end{flalign}
by inverse Fourier-Laplace transforms, one obtains the evolution equation 
\begin{flalign}
\begin{split}
\rho(x,t)=&\rho_0(x)\Psi(t)+\delta(x)\int_{0}^{t}\varphi_s(t')dt' 
+\int_0^t \int_{-\infty}^{\infty}\rho(x-x',t-\tau)\Lambda(x')\varphi_d(\tau)dx' d\tau.
\label{}
\end{split}
\end{flalign}

From Eq. (\ref{ML-eq0}) we also obtain with changeling form
\begin{flalign}
\begin{split}
u\rho(k,u)-\rho_0(k)=\frac{\varphi_d(u)}{\Psi(u)}\rho(k,u)(\Lambda(k)-1)
-\frac{\varphi_s(u)}{\Psi(u)}\rho(k,u)+\frac{\varphi_s(u)}{u\Psi(u)}.
\label{master-eq0}
\end{split}
\end{flalign}
Then with inverse Fourier-Laplace transforms the governing equation with stochastic resetting is given by
\begin{flalign}
\begin{split}
\frac{\partial\rho(x,t)}{\partial t}=\int_0^t M_s(\tau)[\delta(x)-\rho(x,t-\tau)] d\tau
 +\int_0^t \int_{-\infty}^{\infty}M_d(\tau)[\rho(x-x',t-\tau)\Lambda(x')-\rho(x,t-\tau)]dx' d\tau,
\label{}
\end{split}
\end{flalign}
in which the jump and reset waiting times can have arbitrary
distributions, here the function  $M_d(t)$ and $M_s(t)$  denote the memory kernels for the particles diffusion and reset, which  are defined in Laplace space as follows,
\begin{flalign}
\begin{split}
M_d(u)=\frac{\varphi_d(u)}{\Psi(u)},~~~~M_s(u)=\frac{\varphi_s(u)}{\Psi(u)}.
\label{memory kernels-eq0}
\end{split}
\end{flalign}

\subsection{ Jump WTD, reset WTD and Markovian reset}
\label{Markovian reset}
Now we recall the exponential PDF given by  
\begin{flalign}
\begin{split}\psi(t)=\lambda\exp{(-\lambda  t)},  
\label{exp-wait-0}
\end{split}
\end{flalign}
which can be considered as the WTD for jump or reset, that in Laplace space reads
$\psi(u)= \frac{\lambda}{\lambda + u}$ \cite{AE1954},
here $\lambda$ is the average rate of jump or reset events.

For the exponent $\alpha \in (0, 1)$, the power-law PDF is given by
\begin{flalign}
\begin{split} \psi(t)=\tau^{-1}t^{\alpha-1}E_{\alpha,\alpha}(-\tau^{-1}t^ {\alpha}), 
\label{power law WTD-0} 
\end{split}
\end{flalign}
here the generalized two-parameter Mittag-Leffer function \cite{B1953, P1998, GKMR2014} is 
$ E_{ \alpha,\beta}(z)=\sum_{n=0}^{\infty}\frac{z^n}{\Gamma(\alpha n+ \beta)} \label{power law WTD-1}$, its Laplace transform for $\{\alpha, \beta, a\} >0$ reads  \cite {SCK2015} \begin{flalign}
\begin{split} \mathcal{L}\big[t^{\beta-1}E_{\alpha,\beta}(-a t^ {\alpha })\big] =\frac{u^{\alpha-\beta}}{u^{\alpha}+a}, \label{eq-inv-lapl-ml-2-par} \end{split}
\end{flalign}  where $\Gamma (x)$ is the Euler Gamma-function.
We have then the power-law PDF (\ref {power law WTD-0}) in Laplace space
\begin{flalign}
\begin{split}
\psi(u)=\frac{1}{1+\tau u^\alpha}.
\label{reset-PowerLaw-eq-1}
\end{split}
\end{flalign}

In the following text, we consider four typical cases of renewal waiting times corresponding to the different types of WTDs for jump and reset.
When the jump WTD and reset WTD are both following as the exponential PDFs with different parameters $\lambda_d$ and $\lambda_s$ respectively as case one, making use of the exponential PDF (\ref{exp-wait-0}),
from Eqs. (\ref{jump-WTD-eq0}) and (\ref{reset-WTD-eq0}) we then get the PDFs for the events $\tau=\tau_d$ and $\tau=\tau_s$ occurring
\begin{flalign}
\begin{split}
\varphi_{d,s}(t)=\lambda_{d,s}\exp{[-(\lambda_d +\lambda_s) t]},
\label{}
\end{split}
\end{flalign}
which in Laplace space read
\begin{flalign}
\begin{split}
\varphi_{d,s}(u)= \frac{\lambda_{d,s}}{\lambda_d +\lambda_s + u}.
\label{exponential-eq1}
\end{split}
\end{flalign}
Further from Eq. (\ref{memory kernels-eq0})
we can obtain the memory kernels in governing equation
\begin{flalign}
\begin{split}
M_{d,s}(u)=\lambda_{d,s},
\label{exponential-MK-eq0}
\end{split}
\end{flalign}
where we have employed Eq. (\ref{survival-eq3}).

If we consider the power-law jump WTD with the exponent $\alpha$ and the parameter $\tau_d$, and the exponential reset WTD with the parameter $\lambda_s$ as case two.
According to the Eq. (\ref{reset-PowerLaw-eq-1}), the Eq. (\ref{jump-WTD-eq0}) in Laplace space becomes 
\begin{flalign}
\begin{split}\varphi_d(u)=\frac{1}{1+\tau_d (u+\lambda_s)^\alpha},
\label{PL-E-eq0}
\end{split}
\end{flalign}
and the Eq. (\ref{reset-WTD-eq0}) in Laplace space reads
\begin{flalign}
\begin{split}\varphi_s(u)=\frac{\lambda_s\tau_d (u+\lambda_s)^{\alpha-1}}{1+\tau_d (u+\lambda_s)^\alpha}.
\label{PL-eq1}
\end{split}
\end{flalign}
Then the memory kernels (\ref{memory kernels-eq0}) in governing equation can be obtained in the forms 
\begin{flalign}
\begin{split}
M_d(u)=\tau^{-1}_d(u+\lambda_s)^{1-\alpha},
\label{PL-E-MK-eq0}
\end{split}
\end{flalign}
and
\begin{flalign}
\begin{split}
M_s(u)=\lambda_s.
\label{PL-E-MK-eq1}
\end{split}
\end{flalign}

In the third case, for the exponential jump WTD with the parameter $\lambda_d$  and the power-law reset WTD with the exponent $\alpha$ and the parameter $\tau_s$, the PDFs  for the renewal events $\tau=\tau_d$ and $\tau=\tau_s$ occurring behave as in Laplace space
\begin{flalign}
\begin{split}\varphi_d(u)=\frac{\lambda_d\tau_s (u+\lambda_d)^{\alpha-1}}{1+\tau_s (u+\lambda_d)^\alpha}
\label{E-PL-eq0}
\end{split}
\end{flalign}
and
\begin{flalign}
\begin{split}\varphi_s(u)=\frac{1}{1+\tau_s (u+\lambda_d)^\alpha}.
\label{}
\end{split}
\end{flalign}
In this situation the memory kernels
take the forms
\begin{flalign}
\begin{split}
M_d(u)=\lambda_d
\label{E-PL-MK-eq0}
\end{split}
\end{flalign}
and
\begin{flalign}
\begin{split}
M_s(u)=\tau^{-1}_s(u+\lambda_d)^{1-\alpha}.
\label{E-PL-MK-eq1}
\end{split}
\end{flalign}

To study the behaviors of intermittent CTRW with power-law jump and reset WTDs, we can consider 
the standard Pareto-type distributions of the form  $\psi_{d, s} (t) = { \alpha_{d, s} \tau _0 ^{ \alpha_{d, s} }} / {t ^ {1 + \alpha_{d, s} }} $ with $\alpha _{d, s} \in (0, 1)$ for $t> \tau _0$ as the jump and reset WTDs, respectively, at long times \cite{HC2018}. The main reason for this is that at long times (i.e., when $u \to 0$), the standard WTDs of the Pareto type and the Mittag-Leffler functions exhibit highly similar asymptotic expressions in Laplace space \cite {HC2018}, $\psi_{d, s} (u) \simeq 1-\tau_{d, s} u^{ \alpha_{d, s}}\simeq\frac{1}{1+\tau_{d, s} u^{ \alpha_{d, s}}}$ with $\tau_{d, s}=\Gamma(1-\alpha_{d, s})\tau_0^{ \alpha_{d, s}}$. Note that in the following numerical simulation, $\tau_0=0.001$.  Therefore, one can find that in Laplace space
\begin{flalign}
\begin{split}\varphi_{d, s}(u)\simeq\frac{\alpha_{d, s}}{(\alpha_d+\alpha_s)\left(1+\tau_{ds}u^{\alpha_d+\alpha_s}\right)}
\label{PL-PL-eq0}
\end{split}
\end{flalign}
with $\tau_{ds}=\Gamma(1-\alpha_d-\alpha_s)\tau_0^{\alpha_d+\alpha_s}$ for $0<\alpha_d+\alpha_s<1$. Note that in the subsequent discussion of power-law jump and reset WTDs case, the parameters $\alpha_{d, s}$ must satisfy this special condition.
With applying the relation (\ref{survival-eq3}) 
the memory kernels take the forms
\begin{flalign}
\begin{split}
M_{d, s}(u)\simeq\frac{\alpha_{d, s}u^{1-\alpha_d-\alpha_s}}{\tau_{ds}(\alpha_d+\alpha_s)}.
\label{PL-PL-MK-eq0}
\end{split}
\end{flalign}

We focus now on the Markovian resetting for intermittent CTRW, i.e., the reset WTD is following as the exponential PDF with parameter $\lambda_s$ and the jump WTD has any shape of PDF $\psi_d(t)$. In this case the Eq. (\ref{jump-WTD-eq0}) can be written then as
\begin{flalign}
\begin{split}\varphi_d(t)=\exp{(-\lambda_s  t)}\psi_d(t)
\label{}
\end{split}
\end{flalign}
with its Laplace transform
\begin{flalign}
\begin{split}\varphi_d(u)=\psi_d(u+\lambda_s).
\label{Markovian-eq1}
\end{split}
\end{flalign}
And the Eq. (\ref{reset-WTD-eq0}) becomes
\begin{flalign}
\begin{split}\varphi_s(t)=\lambda_s\exp{(-\lambda_s  t)}\Psi_d(t),
\label{}
\end{split}
\end{flalign}
that in Laplace space reads
\begin{flalign}
\begin{split}\varphi_s(u)=\lambda_s\Psi_d(u+\lambda_s).
\label{Markovian-eq2}
\end{split}
\end{flalign}
The survival probability (\ref{survival-eq0}) of the particles not moving away in $[0,t]$ can be found as
\begin{flalign}
\begin{split}
\Psi(t)=\exp{(-\lambda_s  t)}\Psi_d(t),
\label{jump-eq15}
\end{split}
\end{flalign}
its Laplace transform is given by
\begin{flalign}
\begin{split}
\Psi(u)=\Psi_d(u+\lambda_s).
\label{Markovian-eq3}
\end{split}
\end{flalign}

Inserting the Eqs. (\ref{Markovian-eq1}), (\ref{Markovian-eq2}) and (\ref{Markovian-eq3}) into Eq. (\ref{ML-eq0}) and employing the Eq. (\ref{ML-eq01}) we can get in real space
\begin{flalign}
\begin{split}
\rho(x,u|x_0)=\rho_0(x,u+\lambda_s|x_0)+\frac{\lambda_s}{u}\rho_0(x,u+\lambda_s|0),
\label{Markovian-eq4}
\end{split}
\end{flalign}
where we denote the PDFs of the particles for CTRWs with and without stochastic resetting on the initial condition $\rho_0(x)=\delta(x-x_0)$ by $\rho(x,u|x_0)$ and $\rho_0(x,u|x_0)$ respectively in Laplace space.
That can be inverted by Laplace as follows:
\begin{flalign}
\begin{split}
\rho(x,t
|x_0)=\exp{(-\lambda_s t)}\rho_0(x,t|x_0)+\int_0^t \lambda_s\exp{(-\lambda_s \tau)}\rho_0(x,\tau|0)d\tau,
\label{jump-eq32}
\end{split}
\end{flalign}
the Markovian resetting for intermittent CTRWs can be found as corresponding to the first model of subdiffusion with stochastic resetting in Ref. \cite{KG2019}.

\subsection{Probability density functions in Fourier space}
\label{PDF-Fourier space}
With the initial condition $\rho_0(x)=\delta(x)$, that is $\rho_0(k)=1$, and using Eq. (\ref{survival-eq3}) the Eq. (\ref{ML-eq0}) reduces to  
\begin{flalign}
\begin{split}
\rho(k,u)=\frac{\frac{1}{u}(1-\varphi_d(u))}{1-\Lambda(k)\varphi_d(u)}.
\label{ML-eq11}
\end{split}
\end{flalign}

In the case of the exponential jump and reset WTDs,
inserting the Eq. (\ref{exponential-eq1}) into Eq. (\ref{ML-eq11}) one can get
\begin{flalign}
\begin{split}
\rho(k,u)&=\frac{u+\lambda_s}{u(u +\lambda_s+\lambda_d(1-\Lambda(k)))}\\
&=\frac{\lambda_s}{u(\lambda_s+\lambda_d(1-\Lambda(k)))}+\frac{\lambda_d(1-\Lambda(k))}{(u +\lambda_s+\lambda_d(1-\Lambda(k)))(\lambda_s+\lambda_d(1-\Lambda(k)))},
\label{}
\end{split}
\end{flalign}
which can be inverted back to the Fourier space in time to obtain
\begin{flalign}
\begin{split}
\rho(k,t)&=\frac{\lambda_s}{\lambda_s+\lambda_d(1-\Lambda(k))}[1+
\frac{\lambda_d}{\lambda_s}(1-\Lambda(k))\exp{(-(\lambda_s+\lambda_d(1-\Lambda(k)))t)}].
\label{exp-PDF-eq0}
\end{split}
\end{flalign}
So one can find a temporal exponential relaxation towards a
stationary state in this expression.

Considering again the power-law jump WTD and the exponential reset WTD, and using the expansion (\ref{PL-E-eq0}) from (\ref{ML-eq11}) we have 
\begin{flalign}
\begin{split}
\rho(k,u)&=\frac{ (u+\lambda_s)^\alpha}{u((u+\lambda_s)^\alpha+\tau_d^{-1}(1-\Lambda(k)))}\\
&=\frac{1}{u}-\frac{ \tau_d^{-1}(1-\Lambda(k))}{u((u+\lambda_s)^\alpha+\tau_d^{-1}(1-\Lambda(k)))},
\label{}
\end{split}
\end{flalign}
with employing the relation of Eq. (\ref{eq-inv-lapl-ml-2-par}), the PDF of the particles in Fourier space then reads 
\begin{flalign}
\begin{split}
\rho(k,t)&=1-\tau_d^{-1}(1-\Lambda(k))\int_0^t \tau^{\alpha-1}E_{\alpha,\alpha}(-\tau_d^{-1}(1-\Lambda(k))\tau^{\alpha})\exp{(-\lambda_s \tau)} d\tau.
\label{}
\end{split}
\end{flalign}

For the exponential jump WTD and the power-law reset WTD, in the
limit $u\rightarrow 0$ in the Laplace space (which is equivalent to the large time limit $t\rightarrow \infty$) employing the Eq. (\ref{E-PL-eq0}) the expansion (\ref{ML-eq11}) can be written approximately as
\begin{flalign}
\begin{split}
\rho(k,u)&=\frac{(u+\lambda_d)^{1-\alpha}+\tau_s u}{u[(u+\lambda_d)^{1-\alpha}+\tau_s (u+\lambda_d)-\lambda_d\tau_s\Lambda(k)]}\\
&\simeq \frac{1}{u}-\frac{ \lambda_d\tau_s(1-\Lambda(k))}{u[(u+\lambda_d)^{1-\alpha}+\lambda_d\tau_s(1-\Lambda(k))]}.
\label{}
\end{split}
\end{flalign}
In the large time limit $t\rightarrow \infty$ the PDF of the particles in Fourier space has an asymptotic expansion
\begin{flalign}
\begin{split}
\rho(k,t)&\simeq 1-\lambda_d\tau_s(1-\Lambda(k))\int_0^t \tau^{-\alpha}E_{1-\alpha,1-\alpha}(-\lambda_d\tau_s(1-\Lambda(k))\tau^{1-\alpha})\exp{(-\lambda_d \tau)} d\tau.
\label{}
\end{split}
\end{flalign}

In the case of power-law jump and reset WTDs, the propagator (\ref{ML-eq11}) satisfies the following approximate relation:
\begin{flalign}
\begin{split}
\rho(k,u)\simeq \frac{1}{u}-\frac{ \alpha_d(1-\Lambda(k))}{u[(\alpha_d+\alpha_s)\tau_{ds}u^{\alpha_d+\alpha_s}+\alpha_s+\alpha_d(1-\Lambda(k))]}.
\label{}
\end{split}
\end{flalign}

In the large time limit $t\rightarrow \infty$ the PDF of the particles in Fourier space now has an approximate expansion
\begin{flalign}
\begin{split}
\rho(k,t)&\simeq 1-\frac{\alpha_d(1-\Lambda(k))}{(\alpha_d+\alpha_s)\tau_{ds}}\int_0^t \tau^{\alpha_d+\alpha_s-1}E_{\alpha_d+\alpha_s,\alpha_d+\alpha_s}\left(-\frac{\alpha_s+\alpha_d(1-\Lambda(k))}{(\alpha_d+\alpha_s)\tau_{ds}}\tau^{\alpha_d+\alpha_s}\right) d\tau
\label{exp-wait0}
\end{split}
\end{flalign}
 for $0<\alpha_d+\alpha_s<1$.

\subsection{Stationary state}
\label{Stationary state}
 With Eq. (\ref{ML-eq11}) on the initial condition of $\rho_0(x)=\delta(x)$ we can obtain the shape of the stationary state in Fourier space 
\begin{flalign}
\begin{split}
\lim_{t\rightarrow \infty}\rho(k, t)&=\lim_{u\rightarrow 0}u\rho(k,u)=\frac{\gamma_{\varphi_d}}{1-(1-\gamma_{\varphi_d})\Lambda(k)},
\label{SS-eq0}
\end{split}
\end{flalign}
where we define $\gamma_{\varphi_d}$ as the stationary reset probability, i.e., $\gamma_{\varphi_d}\equiv 1-\varphi_d(u=0)$. 
It can be shown that the condition for the existence of a stationary non-equilibrium state is: $\varphi_d(u=0)<1$, which is clearly satisfied by the renewal-reset mechanism.
In this way the result of stationary state weakly similar to that previously obtained in Ref. \cite{MC2016}, but with different temporal tracks of particles due to different reset mechanisms. If the PDF
of jump $\Lambda(x)$ is spatially symmetric, the non-equilibrium stationary state distribution in the real space reads
\begin{flalign}
\begin{split}
\lim_{t\rightarrow \infty}\rho(x, t)&=\frac{1}{\pi}\int_0^{\infty}\frac{\gamma_{\varphi_d}\cos(kx)}{1-(1-\gamma_{\varphi_d})\Lambda(k)}dk\\
&=\gamma_{\varphi_d}\delta(x)+\frac{1}{\pi}\gamma_{\varphi_d}(1-\gamma_{\varphi_d})\int_0^{\infty}\frac{\Lambda(k)\cos(kx)}{1-(1-\gamma_{\varphi_d})\Lambda(k)}dk.
\label{stationary-eq0}
\end{split}
\end{flalign}
With employing the way of expanding the $\cos(kx)$ in power series in Ref.\cite{MC2016}, one can find the approximated shape of the stationary state close to $x = 0$ from Eq. (\ref{stationary-eq0}) reads
\begin{flalign}
\begin{split}
\lim_{t\rightarrow \infty}\rho(x, t)\simeq a_0-a_1 x^2
\label{stationary-eq1}
\end{split}
\end{flalign}
for $x\rightarrow0$, where the factors $a_0$ and $a_1$ are given by
\begin{flalign}
\begin{split}
&a_0= \gamma_{\varphi_d}\delta(x)+\frac{1}{\pi}\gamma_{\varphi_d}(1-\gamma_{\varphi_d})\int_0^{\infty}\frac{\Lambda(k)}{1-(1-\gamma_{\varphi_d})\Lambda(k)}dk,\\
&a_1=\frac{1}{2\pi}\gamma_{\varphi_d}(1-\gamma_{\varphi_d})\int_0^{\infty}\frac{k^2\Lambda(k)}{1-(1-\gamma_{\varphi_d})\Lambda(k)}dk.
\end{split}
\end{flalign}
Note that, unlike Ref. \cite{MC2016}, the stationary reset probability $\gamma_{\varphi_d}$ is not an independent variable, but depends on the jump and reset WTDs.

\section{Examples and some results}
\subsection{Exponential distributed jump length}
\label{Exponential distributed}
We now consider the exponential distributed jump length
\begin{flalign}
\begin{split}
\Lambda(\xi)
=\frac{1}{2\omega}\exp\{-\frac{|\xi|}{\omega}\}
\label{Jump-length-E}
\end{split}
\end{flalign}
with $\omega>0$  for intermittent CTRWs.
Its Fourier transform reads
\begin{flalign}
\begin{split}
\Lambda(k)
=\frac{1}{1+k^2\omega^2}.
\label{lambda-E }
\end{split}
\end{flalign}

Employing Eq. (\ref{lambda-E }) into Eq. (\ref{master-eq0}) one gets
\begin{flalign}
\begin{split}
u\rho(k,u)-\rho_0(k)= &M_d(u)\rho(k,u)\left(\frac{1}{1+k^2\omega^2}-1\right)+M_s(u)[\frac{1}{u}-\rho(k,u)].
\label{}
\end{split}
\end{flalign}
Taking the inverse Laplace-Fourier transforms one then can obtain the governing equation with the exponential
distributed jump length
\begin{flalign}
\begin{split}
\frac{\partial\rho(x,t)}{\partial t}=&\int_0^t M_s(t-\tau)[\delta(x)-\rho(x,\tau)] d\tau +\int_0^t d\tau M_d(t-\tau)\\
&\times\left[\frac{1}{2\omega}\int_{-\infty}^{\infty}\exp\{-\frac{|x-x'|}{\omega}\}\rho(x',\tau)dx'-\rho(x,\tau)\right]
\label{E-master-eq0}
\end{split}
\end{flalign}
for arbitrary jump and
reset WTDs.

Combing Eq. (\ref{lambda-E }) from Eq. (\ref{ML-eq11}) with the initial condition $\rho_0(x)=\delta(x)$ we then
get \begin{flalign}
\begin{split}
\rho(k,u)=\frac{1}{u}(1-\varphi_d(u))+\frac{\frac{1}{u}(1-\varphi_d(u))\varphi_d(u)}{1-\varphi_d(u)+k^2\omega^2}.
\label{balance-eq33}
\end{split}
\end{flalign}
The PDF of particles in Laplace space can be inverted exactly by Fourier transform as follows:
\begin{flalign}
\begin{split}
\rho(x,u)=&\frac{1}{2\omega u}\varphi_d(u)\sqrt{1-\varphi_d(u)}\exp\left\{-\frac{1}{\omega }\sqrt{1-\varphi_d(u)}\big|x\big|\right\}+\frac{1}{u}(1-\varphi_d(u))\delta(x).
\label{E-PDF-eq0}
\end{split}
\end{flalign}

For the exponential distributed jump length, by solving for the first and second order derivatives of
Eq. (\ref{ML-eq11}) with respect to the variable $k$ and employing Eq. (\ref{lambda-E }) the first and second moments in Laplace space can be obtained as follows:
\begin{flalign}
\begin{split}
&\langle x(u)\rangle =i\frac{\partial}{\partial k}\rho(k,u)|_{k=0}= 0,\\
&\langle x^2(u)\rangle =i^2\frac{\partial^2}{\partial k^2}\rho(k,u)|_{k=0}=\frac{2\omega^2\varphi_d(u)}{u(1-\varphi_d(u))}.
\label{E-MSD-eq0}
\end{split}
\end{flalign}
Where the first moment satisfies $\langle x(t)\rangle =0$ due to the spatially symmetric jump length.

To study a search process, the 
MFAT serves as an important statistical indicator to quantify search effectiveness. Generally speaking, diffusion processes with reset lead to a finite MFAT to any fixed position \cite{KG2019}. We now present the derivation of the first-passage time statistics, which will help elucidate the characteristics of  intermittent CTRW with stochastic resetting.
Let the process begin at $t_0 = 0$ with $x_0 = 0$. The distribution of the FAT at the target position $x$ (denoted as $\chi$ below) relates to the free propagator $\rho$ (without absorbing boundaries) as follows \cite{KG2019}:
\begin{flalign}
\begin{split}
\rho(x,t|x_0)=\int_0^t d\tau \chi(\tau)\rho(x,t-\tau|x).
\label{Eq-FAT1}
\end{split}
\end{flalign}
In Laplace space, equation (\ref{Eq-FAT1}) simplifies to the following equation
\begin{flalign}
\begin{split}
\rho(x,u|0)=\chi(u)\rho(x,u|x).
\label{Eq-FAT2}
\end{split}
\end{flalign}
By applying the propagation (\ref{ML-eq0}) of intermittent CTRW and the general relationship (\ref{Eq-FAT2}) satisfied by the first-arrival time (FAT), we can derive the important expression in Laplace space for FAT in intermittent CTRW with stochastic resetting through several steps of algebraic operations,
\begin{flalign}
\begin{split}
\chi(u)=\frac{1-\varphi_d(u)}{u\Psi(u)\left[1+\frac{2\omega\sqrt{1-\varphi_d(u)}}{\varphi_d(u)}\right]\exp{(\frac{\sqrt{1-\varphi_d(u)}}{\omega}\big|x_0\big|)}+\varphi_s(u)},
\label{Eq-FAT3}
\end{split}
\end{flalign}
which can be used to solve for MFAT through its first derivative.

In case of the exponential jump and
reset WTDs, combing the memory kernels (\ref{exponential-MK-eq0}) the Eq. (\ref{E-master-eq0})
reduces to
\begin{flalign}
\begin{split}
\frac{\partial\rho(x,t)}{\partial t}=&\frac{\lambda_d}{2\omega}\int_{-\infty}^{\infty}\exp\{-\frac{|x-x'|}{\omega}\}\rho(x',t)dx'-(\lambda_d+\lambda_s)\rho(x,t)+\lambda_s\delta(x).
\label{}
\end{split}
\end{flalign}

With Eq. (\ref{exponential-eq1}) from Eq. (\ref{E-PDF-eq0}) the shape of the stationary state can be derived as
 follows:
\begin{flalign}
\begin{split}
\lim_{t\rightarrow \infty}\rho(x, t)=&\lim_{u\rightarrow 0}u\rho(x,u)\\
=&\frac{1}{2\omega}\frac{\lambda_d}{\lambda_d+\lambda_s}\sqrt{\frac{\lambda_s}{\lambda_d+\lambda_s}}\exp\left\{-\frac{1}{\omega }\sqrt{\frac{\lambda_s}{\lambda_d+\lambda_s}}\big|x\big|\right\}+\frac{\lambda_s}{\lambda_d+\lambda_s}\delta(x).
\label{PDF1-E}
\end{split}
\end{flalign}
And the MSD (the second moment) can also be derived from Eq. (\ref{E-MSD-eq0}) with inverse Laplace transform in this case 
\begin{flalign}
\begin{split}
\langle x^2(t)\rangle = \frac{2\omega^2\lambda_d}{\lambda_s}\left(1-\exp{(-\lambda_s t)}\right).
\label{MSD1-E}
\end{split}
\end{flalign}
It follows from the relation (\ref{MSD1-E}) that, for a constant jump length parameter $\omega$, the jump reset competition is determined by the ratio of their parameters $\lambda_d/\lambda_s$, which specifies the maximum amplitude of the MSD. In fact, this ratio is also the only key parameter that determines the stationary distribution at this time.

The distribution of the FAT at the target position $x_0$ can be simplified by substituting relations (\ref{exponential-eq1}) and (\ref{survival-eq3}) into expression (\ref{Eq-FAT3}),
\begin{flalign}
\begin{split}
\chi(u)=\frac{\lambda_s+u}{u\left[1+\frac{2\omega}{\lambda_d}\sqrt{(\lambda_s+u)(\lambda_d+\lambda_s+u)}\right]\exp{\left(\frac{|x_0|}{\omega}\sqrt{\frac{\lambda_s+u}{\lambda_d+\lambda_s+u}}\right)}+\lambda_s}.
\label{Eq-FAT4}
\end{split}
\end{flalign}
The MFAT value for the random variable $T$ of FAT can be obtained by solving the first derivative of equation (\ref{Eq-FAT4}) at the point $u=0$. 
\begin{flalign}
\begin{split}
<T>=\frac{1}{\lambda_s}\left[\left[1+\frac{2\omega}{\lambda_d}\sqrt{\lambda_s(\lambda_d+\lambda_s)}\right]\exp{\left(\frac{|x_0|}{\omega}\sqrt{\frac{\lambda_s}{\lambda_d+\lambda_s}}\right)}-1\right].
\label{Eq-FAT5}
\end{split}
\end{flalign}
In fact one can adjust the frequency parameter $\lambda_s$ of reset events to minimize the MFAT to improve the effectiveness of random search. If the parameter $\lambda_s$ is considered an independent variable of MFAT function $<T>$, the  minimum value of MFAT exists and can be attained at the stationary point.

\begin{figure}[htbp]
\centering
\begin{tabular}{cc}
\includegraphics[width=0.48\textwidth]{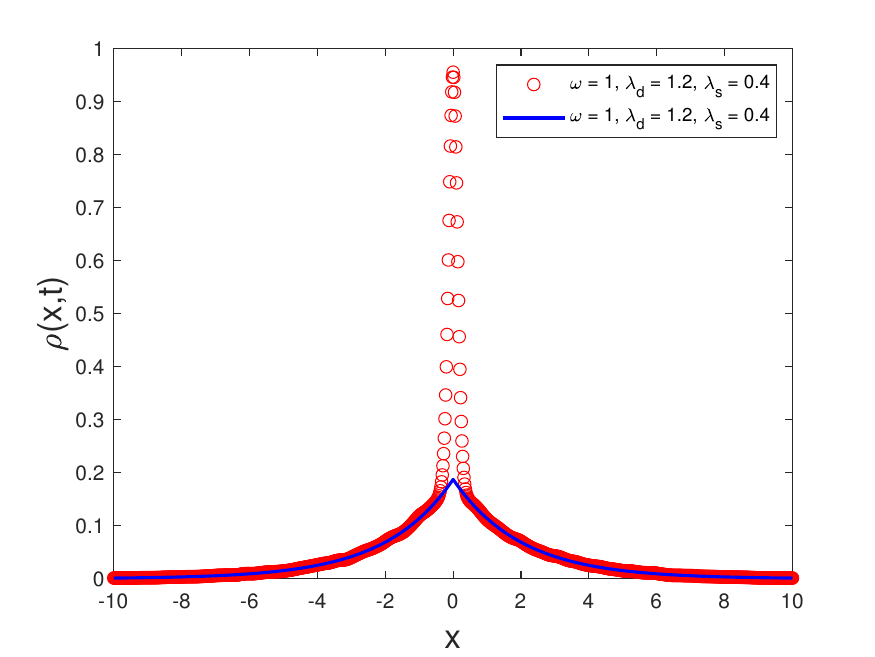} &
\includegraphics[width=0.48\textwidth]{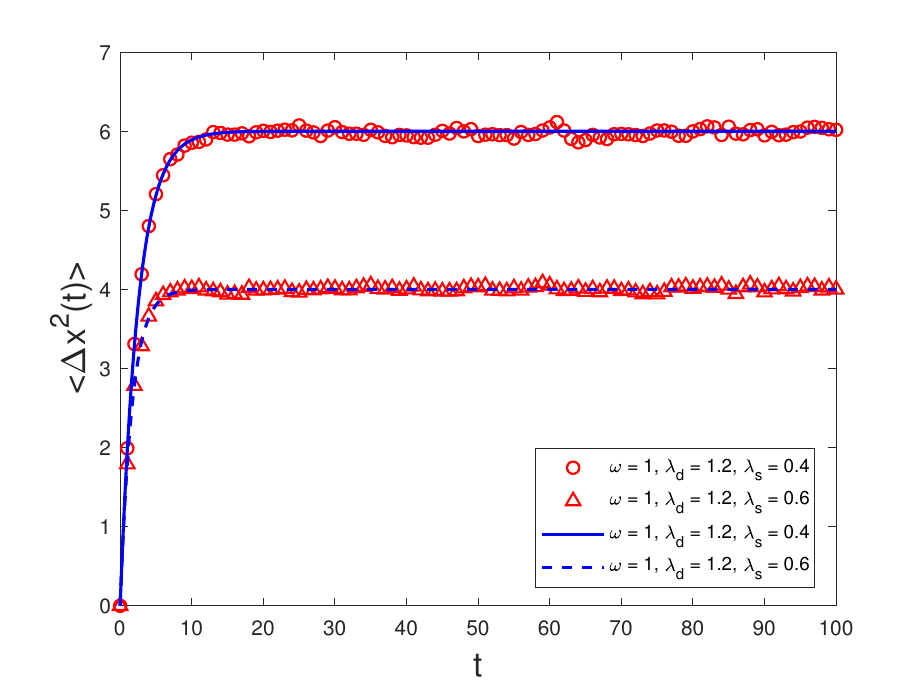} \\
(a) & (b) \\[0.3cm]
\multicolumn{2}{c}{\includegraphics[width=0.48\textwidth]{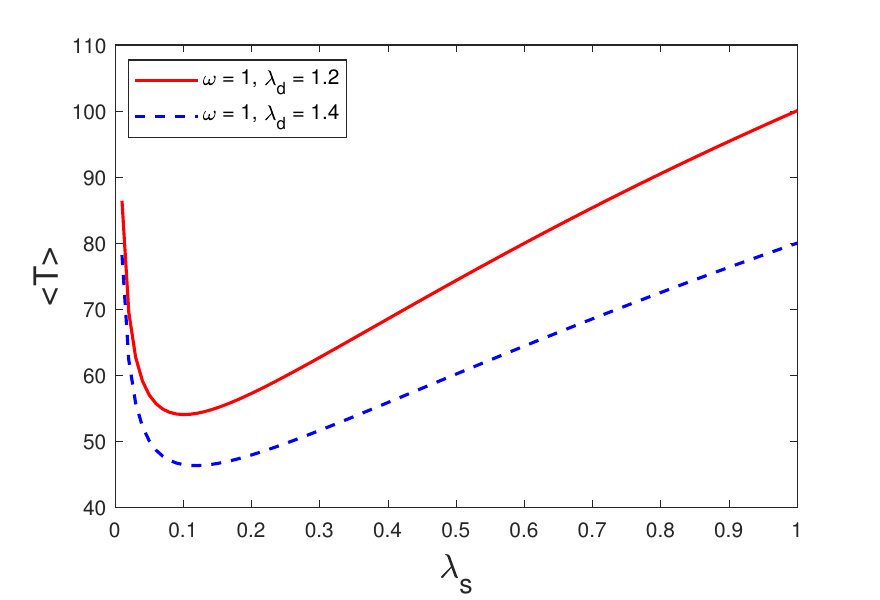}} \\
\multicolumn{2}{c}{(c)}
\end{tabular}
	\caption{ Results of Monte-Carlo simulations (symbols) versus the analytical solutions (solid curves) for the shape (\ref {PDF1-E}) of the stationary state (removed the last item containing the generalized function $\delta(x)$) in panel (a) and the MSDs (\ref {MSD1-E}) in panel (b) for the intermittent CTRWs with exponential distributed jump length (\ref{Jump-length-E}) for the exponential jump and reset WTDs  (\ref {exp-wait-0}). The simulation results  were plotted at the diffusion time $t = 100$ (including simulation for the shape of the stationary state) and averaged over $N=10^5$ trajectories of the particles. In panel (c) the analytical results of MFAT (\ref{Eq-FAT5}) at the target position $x_0=5$ as the parameter $\lambda_s$ varies to show the existence of its minimum value through reset parameter adjustment.
    The values of relevant model parameters are provided in the legend. }
	\label{fig_1ab}
\end{figure}

To demonstrate the accuracy and validity of the intermittent CTRW model under renewal reset mechanism, we have now performed simulations based on jump lengths following the exponential distribution to check the PDF in the stationary state and the MSDs for intermittent CTRWs in Figs. \ref{fig_1ab}(a) and \ref{fig_1ab}(b). These simulation results were consistent with the theoretical analysis of (\ref{PDF1-E}) and (\ref{MSD1-E}) and, at the same time, confirmed the existence of a non-equilibrium stationary state in the new reset mechanism and showed that the diffusivity of intermittent CTRWs decreases with increasing reset frequency as the analysis result of MSD. In Fig. \ref{fig_1ab}(c) we demonstrate the evolution of the analytical value of MFAT (\ref{Eq-FAT5}) as the parameter $\lambda_s$ varies and confirm the existence of its minimum value through reset parameter adjustment.

For the power-law jump and exponential reset WTDs, making use of the memory kernels (\ref{PL-E-MK-eq0}) and (\ref{PL-E-MK-eq1}) the Eq. (\ref{E-master-eq0})
can be written as
\begin{flalign}
\begin{split}
\frac{\partial\rho(x,t)}{\partial t}=&\frac{1}{2\omega \tau_d} e^{-\lambda_s t}{}_{0}D_t^{1-\alpha} \bigg[e^{\lambda_s t}\bigg(\int_{-\infty}^{\infty}\exp\{-\frac{|x-x'|}{\omega}\}\rho(x',t)dx'-2\omega\rho(x,t)\bigg)\bigg]+\lambda_s[\delta(x)-\rho(x,t)].
\label{}
\end{split}
\end{flalign}
Here, the diffusion operator  \cite{HLW2008,SS2007}
\begin{eqnarray}
{e^{-\lambda t}{}_{0}D_t^{1-\alpha} \left(e^{\lambda t} f(t)\right)}=\frac{1}{\Gamma(\alpha)}\bigg(\frac{d
}{dt}\int_{0}^{t}\frac{e^{-\lambda(t-t')}}{(t-t')^{1-\alpha}}f(t')dt'+\lambda\int_{0}^{t}\frac{e^{-\lambda(t-t')}}{(t-t')^{1-\alpha}}f(t')dt'\bigg)
\end{eqnarray}
is equal in Laplace $t\rightarrow u$ space to $(u+\lambda)^{1-\alpha}f(u)$, and when $\lambda=0$ it becomes ${}_{0}D_t^{1-\alpha}f(t)$  which is the  Riemann-Liouville fractional derivative operator \cite{MK2000,MR1993}.

Using Eq. (\ref{PL-E-eq0}) from Eq. (\ref{E-PDF-eq0}) the shape of the stationary state reads
\begin{flalign}
\begin{split}
\lim_{t\rightarrow \infty}\rho(x, t)=&\frac{1}{2\omega}\frac{1}{1+\tau_d \lambda_s^\alpha}\sqrt{\frac{\tau_d \lambda_s^\alpha}{1+\tau_d \lambda_s^\alpha}}\exp\left\{-\frac{1}{\omega }\sqrt{\frac{\tau_d \lambda_s^\alpha}{1+\tau_d \lambda_s^\alpha}}\big|x\big|\right\}+\frac{\tau_d \lambda_s^\alpha}{1+\tau_d \lambda_s^\alpha}\delta(x).
\label{Eq-PDF-E2}
\end{split}
\end{flalign}
In the figures below, we use $\rho_s(x)$ to denote the PDF of particles in the stationary state, that is, $\rho_s(x)=\lim_{t\rightarrow \infty}\rho(x, t)$.

And at this situation from Eq. (\ref{E-MSD-eq0}) the MSD (the second moment) can also be obtained as follows:
\begin{flalign}
\begin{split}
\langle x^2(t)\rangle = \frac{2\omega^2}{\Gamma(\alpha)\tau_d \lambda_s^\alpha} \gamma(\alpha, \lambda_s t),
\label{Eq-MSD-E2}
\end{split}
\end{flalign}
here the incomplete Gamma-function is given by \cite{JD2008}
\begin{flalign}
\begin{split}
\gamma(\alpha, t)=\int_0^t \tau^{\alpha-1}\exp{(-\tau)} d\tau.
\label{}
\end{split}
\end{flalign}

Following the calculation process described above, we can calculate the distribution of the FAT by inserting the relationships (\ref{PL-E-eq0}) and (\ref{survival-eq3}) into the expression (\ref{Eq-FAT3}),
\begin{flalign}
\begin{split}
\chi(u)=\frac{u+\lambda_s}{u\left[1+2\omega\sqrt{\tau_d(u+\lambda_s)^\alpha(1+\tau_d(u+\lambda_s)^\alpha)}\right]\exp{\left(\frac{|x_0|}{\omega}\sqrt{ \frac{\tau_d(u+\lambda_s)^\alpha}{1+\tau_d(u+\lambda_s)^\alpha}}\right)}+\lambda_s},
\label{}
\end{split}
\end{flalign}
and the MFAT can be solved to yield the following form:
\begin{flalign}
\begin{split}
<T>=\frac{1}{\lambda_s}\left[\left[1+2\omega\sqrt{\tau_d \lambda_s^\alpha(1+\tau_d\lambda_s^\alpha)}\right]\exp{\left(\frac{|x_0|}{\omega}\sqrt{ \frac{\tau_d\lambda_s^\alpha}{1+\tau_d\lambda_s^\alpha}}\right)}-1\right].
\label{Eq-MFAT-E2}
\end{split}
\end{flalign}

\begin{figure}[htbp]
\centering
\begin{tabular}{cc}
\includegraphics[width=0.48\textwidth]{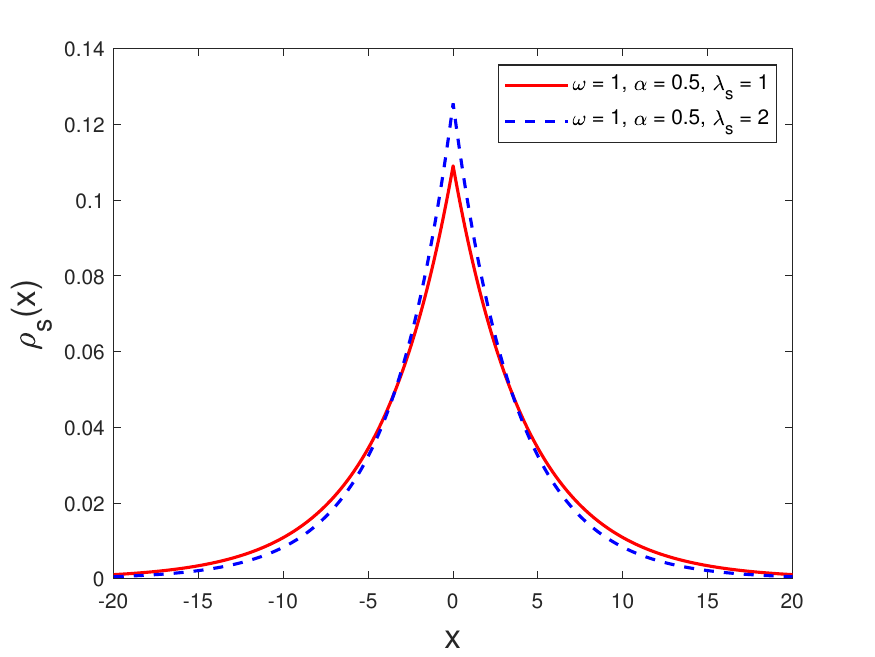} &
\includegraphics[width=0.48\textwidth]{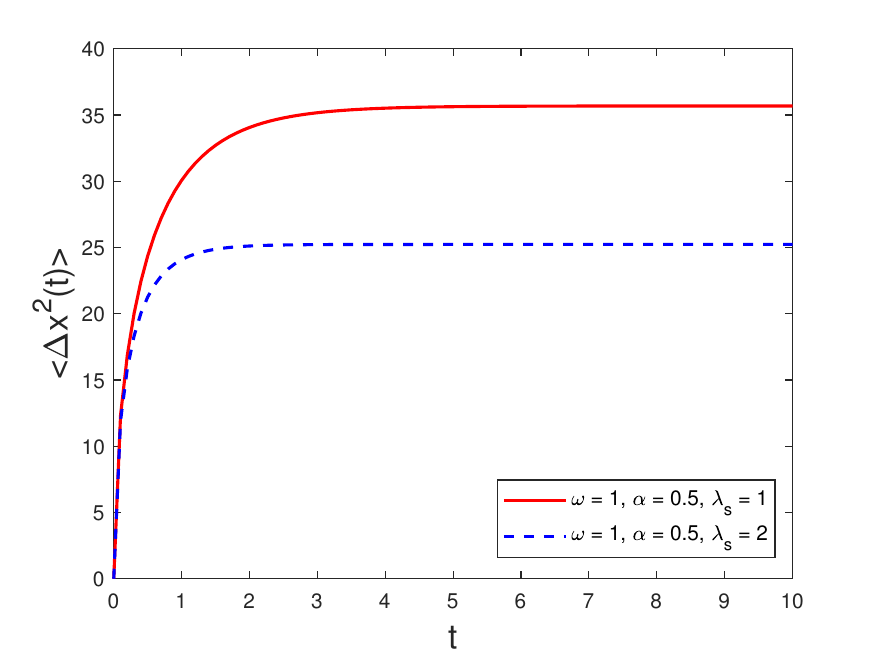} \\
(a) & (b) \\[0.3cm]
\multicolumn{2}{c}{\includegraphics[width=0.48\textwidth]{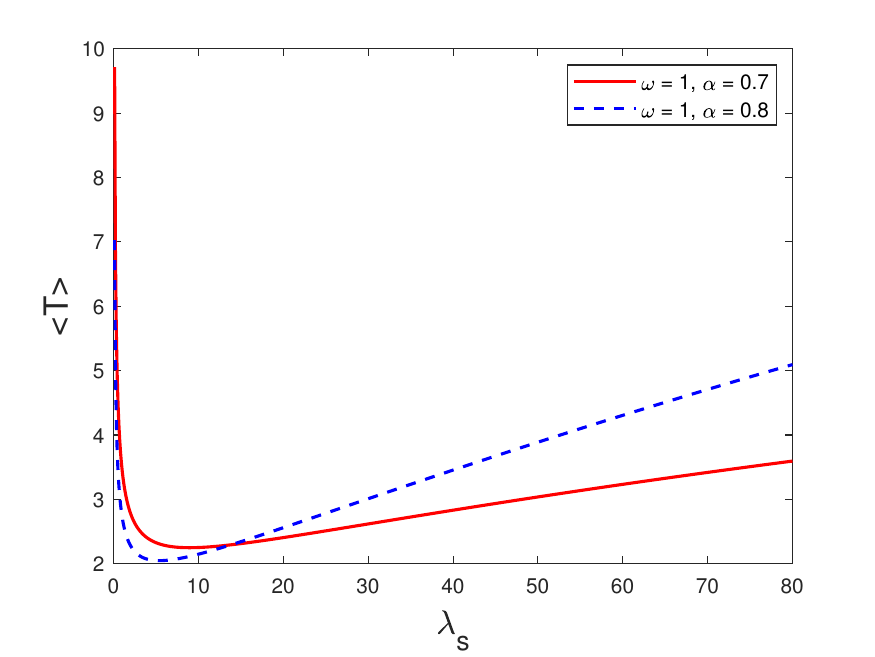}} \\
\multicolumn{2}{c}{(c)}
\end{tabular}
\caption{The analytical solutions for the shape (\ref {Eq-PDF-E2}) of the stationary state (removed the last item containing the generalized function $\delta(x)$) in panel (a) and the MSDs (\ref {Eq-MSD-E2}) in panel (b) for the intermittent CTRWs with power-law jump WTD and exponential reset WTD. In panel (c) the analytical results of MFAT (\ref{Eq-MFAT-E2}) at the target position $x_0=8$ as the reset parameter $\lambda_s$ varies. The values of relevant model parameters are provided in the legend.}
\label{Fig-E2}
\end{figure}

It should be noted that in studying the diffusion behavior of CTRW model, the precise simulation sampling of Mittag-Leffler type (power-law) WTDs remains an unsolved problem. To our knowledge, the currently prevailing approximation, in which simulation sampling is performed by the standard Pareto-type distribution replacement, consistently yields biased results. This distortion persists even when the theoretical and simulated results appear to be nearly overlap or parallel in the representations of the loglog plots. For this reason, this article does not address the simulation of intermittent CTRW for the WTD of Mittag-Leffler type here. Figs. \ref{Fig-E2}(a) and \ref{Fig-E2}(b) show the analysis results for stationary state distributions of particles and the MSD. Compared to the jump WTD of the exponential distribution, the power-law WTD of jump changes the temporal evolution properties of the MSD (\ref{Eq-MSD-E2}), but retains its monotonically increasing character. The distribution type (\ref{Eq-PDF-E2}) of the corresponding stationary state remains unchanged. Figs. \ref{Fig-E2}(a) and \ref{Fig-E2}(b) show that an increase in the reset parameter $\lambda_s$ amplifies the peak of the stationary state distribution and at the same time significantly reduces the second moment of intermittent CTRWs. Fig. \ref{Fig-E2}(c) illustrates the evolution of the finite MFAT (\ref{Eq-MFAT-E2}) as a function of the reset parameter $\lambda_s$ and simultaneously demonstrates the effectiveness of stochastic resetting in improving random search within the CTRW framework with power-law jump WTDs.

For exponential jump and power-law reset WTDs,  using the memory kernels (\ref{E-PL-MK-eq0}) and (\ref{E-PL-MK-eq1}) into the governing equation (\ref{E-master-eq0}) becomes
\begin{flalign}
\begin{split}
\frac{\partial\rho(x,t)}{\partial t}=\tau_s^{-1}e^{-\lambda_d t}{}_{0}D_t^{1-\alpha} \left(e^{\lambda_d t}[\delta(x)-\rho(x,t)]\right)+\frac{\lambda_d}{2\omega}\int_{-\infty}^{\infty}\exp\{-\frac{|x-x'|}{\omega}\}\rho(x',t)dx'-\lambda_d\rho(x,t).
\label{}
\end{split}
\end{flalign}
Employing Eq. (\ref{E-PL-eq0}) into Eq. (\ref{E-PDF-eq0}) we can find the stationary state as follows:
\begin{flalign}
\begin{split}
\lim_{t\rightarrow \infty}\rho(x, t)=\frac{1}{2\omega}\frac{\tau_s \lambda_d^\alpha}{1+\tau_s \lambda_d^\alpha}\sqrt{\frac{1}{1+\tau_s \lambda_d^\alpha}}\exp\left\{-\frac{1}{\omega }\sqrt{\frac{1}{1+\tau_s \lambda_d^\alpha}}\big|x\big|\right\}+\frac{1}{1+\tau_s \lambda_d^\alpha}\delta(x).
\label{Eq-PDF-E3}
\end{split}
\end{flalign}
And using Eq. (\ref{E-PL-eq0}) into the Eq. (\ref{E-MSD-eq0}), with inverse Laplace transform the approximated MSD (the second moment) in the large time limit can be written as follows: 
\begin{flalign}
\begin{split}
\langle x^2(t)\rangle \simeq \frac{2\omega^2\lambda_d^{\alpha} \tau_s}{\Gamma(1-\alpha) } \gamma(1-\alpha, \lambda_d t).
\label{Eq-MSD-E3}
\end{split}
\end{flalign}

In the case of exponential jump and power-law reset WTDs, 
the distribution of the FAT can be determined by inserting the relationships (\ref{E-PL-eq0}) and (\ref{survival-eq3}) into the expression (\ref{Eq-FAT3}),
\begin{flalign}
\begin{split}
\chi(u)=\frac{\tau_su+(u+\lambda_d)^{1-\alpha}}{\tau_suC(u)\exp{\left(\frac{|x_0|}{\omega}\sqrt{\frac{1+\tau_s u(u+\lambda_d)^{\alpha-1}}{1+\tau_s(u+\lambda_d)^{\alpha}}}\right)}+(u+\lambda_d)^{1-\alpha}}, 
\label{}
\end{split}
\end{flalign}
with $C(u)=1+\frac{2\omega}{\lambda_d\tau_s (u+\lambda_d)^{\alpha-1}}\sqrt{(1+\tau_s u(u+\lambda_d)^{\alpha-1})(1+\tau_s(u+\lambda_d)^{\alpha})}$.
Then we find
\begin{flalign}
\begin{split}
<T>=\frac{\tau_s}{\lambda_d^{1-\alpha}}\left[\left(1+\frac{2\omega\sqrt{1+\tau_s\lambda_d^{\alpha}}}{\tau_s \lambda_d^{\alpha}}\right)\exp{\left(\frac{|x_0|}{\omega\sqrt{1+\tau_s\lambda_d^{\alpha}}}\right)}-1\right].
\label{Eq-MFAT-E3}
\end{split}
\end{flalign}

\begin{figure}[htbp]
\centering
\begin{tabular}{cc}
\includegraphics[width=0.48\textwidth]{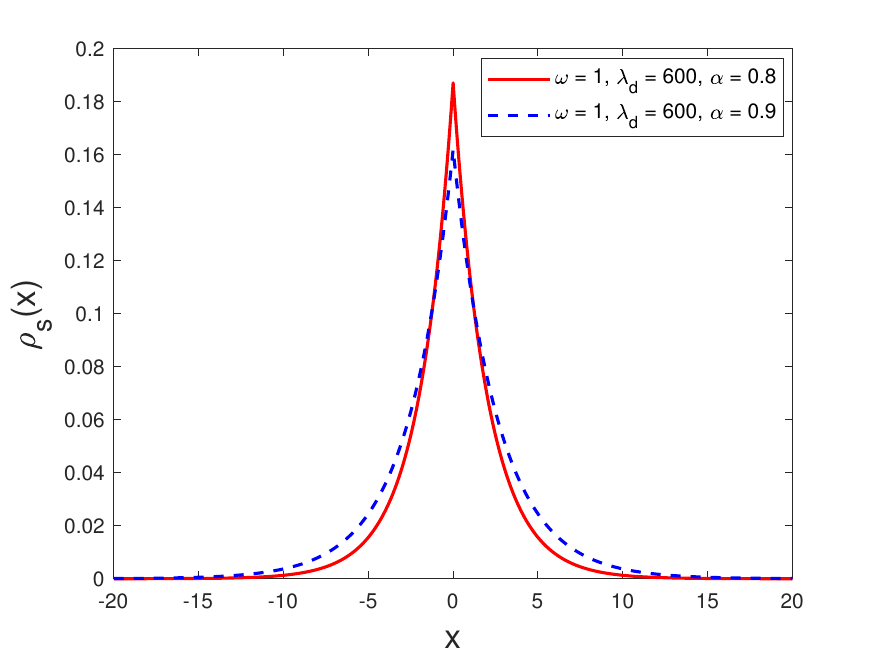} &
\includegraphics[width=0.48\textwidth]{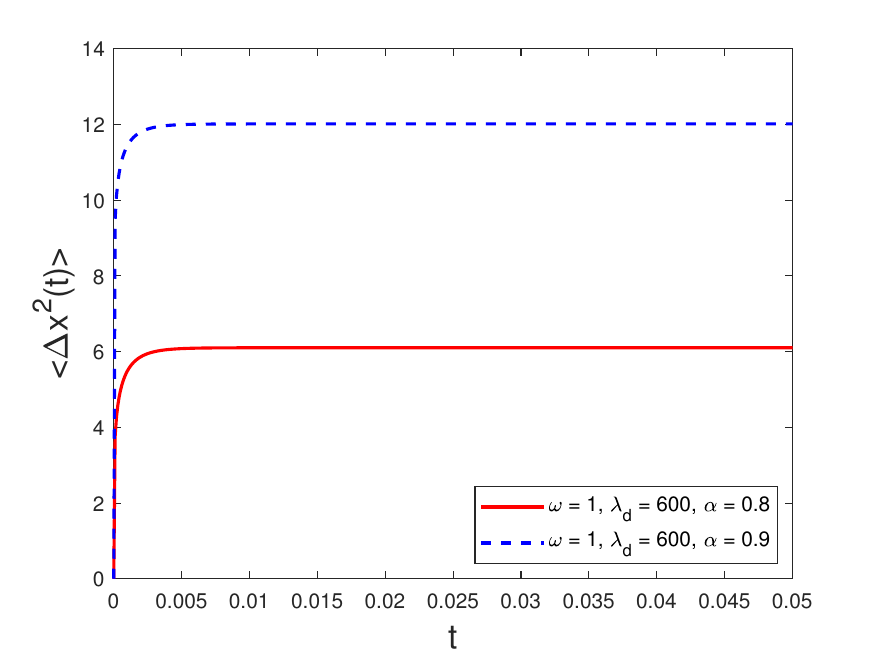} \\
(a) & (b) \\[0.3cm]
\multicolumn{2}{c}{\includegraphics[width=0.48\textwidth]{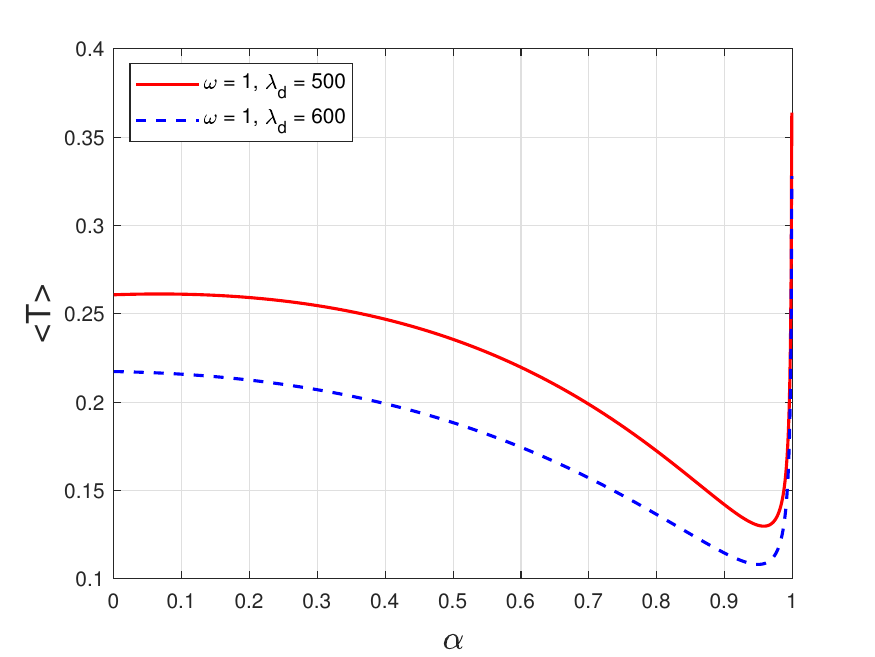}} \\
\multicolumn{2}{c}{(c)}
\end{tabular}
\caption{The distribution in non-equilibrium stationary state (\ref{Eq-PDF-E3})(removed the last item containing the generalized function $\delta(x)$) in panel (a) and the MSDs (\ref{Eq-MSD-E3}) in panel (b) for the intermittent CTRWs with power-law reset WTD. In panel (c) the analytical results of MFAT (\ref{Eq-MFAT-E3}) at the target position $x_0=5$ as the reset power-law distribution exponent $\alpha$ varies. The values of relevant model parameters are provided in the legend.}
\label{Fig-E3}
\end{figure}

In Fig. \ref{Fig-E3}(a), we show the distribution in non-equilibrium stationary state (\ref{Eq-PDF-E3}), which indicates that the peak decreases with increasing reset power-law distribution exponent. Accordingly, in Fig. \ref{Fig-E3}(b), the MSD (\ref{Eq-MSD-E3}) increases significantly but maintains its monotonically increasing trend. Based on the analytical expression for MFAT (\ref{Eq-MFAT-E3}), it can be observed that under renewal reset mechanism, the reset WTD of the power-law distribution still yields a finite MFAT. Interestingly, MFAT shows a development with multiple peaks and multiple stationary points when the reset power-aw distribution exponent changes. In the analytical plot in Fig. \ref{Fig-E3}(c) for a given set of parameters, the reset power-law distribution exponent close to $0.95$ minimizes the MFAT, thus achieving optimal search-ability.

As discussed earlier, we can consider the standard Pareto-type distributions as the jump and reset WTDs to study the behaviors of intermittent CTRWs with stochastic resetting. Substituting the memory kernels (\ref{PL-PL-MK-eq0}) into the governing equation (\ref{E-master-eq0}) yields
\begin{flalign}
\begin{split}
\frac{\partial\rho(x,t)}{\partial t}\simeq\frac{1}{\tau_{ds}(\alpha_d+\alpha_s)}{}_{0}D_t^{1-\alpha_d-\alpha_s}\bigg[\frac{\alpha_d}{2\omega}\int_{-\infty}^{\infty}\exp\{-\frac{|x-x'|}{\omega}\}\rho(x',t)dx'-(\alpha_d+\alpha_s)\rho(x,t)+\alpha_s\delta(x)\bigg]
\label{PL-PL-GOV-E1}
\end{split}
\end{flalign}
for $0<\alpha_d+\alpha_s<1$.

We can find the distribution of stationary state by employing Eq. (\ref{PL-PL-eq0}) into Eq. (\ref{E-PDF-eq0}),
\begin{flalign}
\begin{split}
\lim_{t\rightarrow \infty}\rho(x, t)=\frac{1}{2\omega}\frac{\alpha_d}{\alpha_d+\alpha_s}\sqrt{\frac{\alpha_s}{\alpha_d+\alpha_s}}\exp\left\{-\frac{1}{\omega }\sqrt{\frac{\alpha_s}{\alpha_d+\alpha_s}}\big|x\big|\right\}+\frac{\alpha_s}{\alpha_d+\alpha_s}\delta(x).
\label{PL-PL-PDF-E1}
\end{split}
\end{flalign}
And the approximated MSD (the second moment) in the large time limit can also be obtained by using Eq. (\ref{PL-PL-eq0}) into the Eq. (\ref{E-MSD-eq0}) with inverse Laplace transform,
\begin{flalign}
\begin{split}
\langle x^2(t)\rangle \simeq \frac{2\omega^2\alpha_d}{\tau_{ds}(\alpha_d+\alpha_s)} t^{\alpha_d+\alpha_s}E_{\alpha_d+\alpha_s,\alpha_d+\alpha_s+1}\left(-\frac{\alpha_st^{\alpha_d+\alpha_s}}{\tau_{ds}(\alpha_d+\alpha_s)} \right).
\label{PL-PL-MSD-E1}
\end{split}
\end{flalign}

For power-law jump and reset WTDs, one can find
the distribution of the FAT in Laplace space by inserting the relationships (\ref{PL-PL-eq0}) and (\ref{survival-eq3}) into the expression (\ref{Eq-FAT3}),
\begin{flalign}
\begin{split}
\chi(u)=\frac{(\alpha_d+\alpha_s)\tau_{ds}u^{\alpha_d+\alpha_s}+\alpha_s}{(\alpha_d+\alpha_s)\tau_{ds}u^{\alpha_d+\alpha_s}C'(u)\exp{\left(\frac{|x_0|}{\omega}\sqrt{\frac{(\alpha_d+\alpha_s)\tau_{ds}u^{\alpha_d+\alpha_s}+\alpha_s}{(\alpha_d+\alpha_s)(1+\tau_{ds}u^{\alpha_d+\alpha_s})}}\right)}+\alpha_s},
\label{Eq-FAT-E4}
\end{split}
\end{flalign}
with $C'(u)=1+\frac{2\omega}{\alpha_d}\sqrt{(\alpha_d+\alpha_s)(1+\tau_{ds}u^{\alpha_d+\alpha_s})[(\alpha_d+\alpha_s)\tau_{ds}u^{\alpha_d+\alpha_s}+\alpha_s]}$.
Unfortunately, by calculating the first derivative of the FAT (\ref{Eq-FAT-E4}) in Laplace space at $u=0$, it is found that the MFAT diverges, that is
\begin{flalign}
\begin{split}
<T>=\infty.
\label{}
\end{split}
\end{flalign}
In our opinion, the divergence characteristic of MFAT arises from the fact that both the power-law jump and reset WTDs have infinite means, which means that the system's renewal time is extremely large.  Consequently, the time required for searchers (predators or random walkers) to reach the target point becomes infinitely long.

\begin{figure}[htbp]
\centering
\begin{tabular}{cc}
\includegraphics[width=0.48\textwidth]{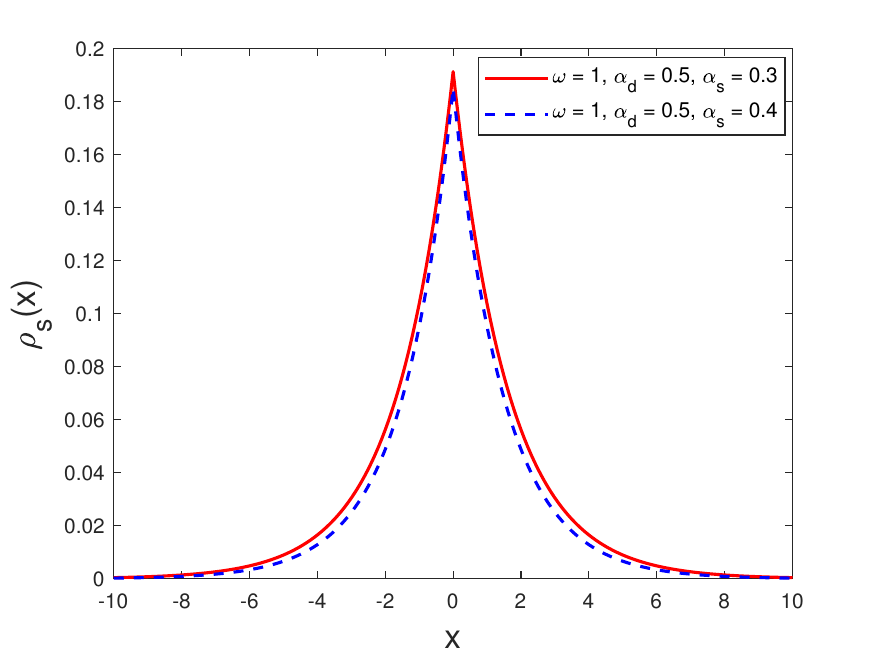} &
\includegraphics[width=0.48\textwidth]{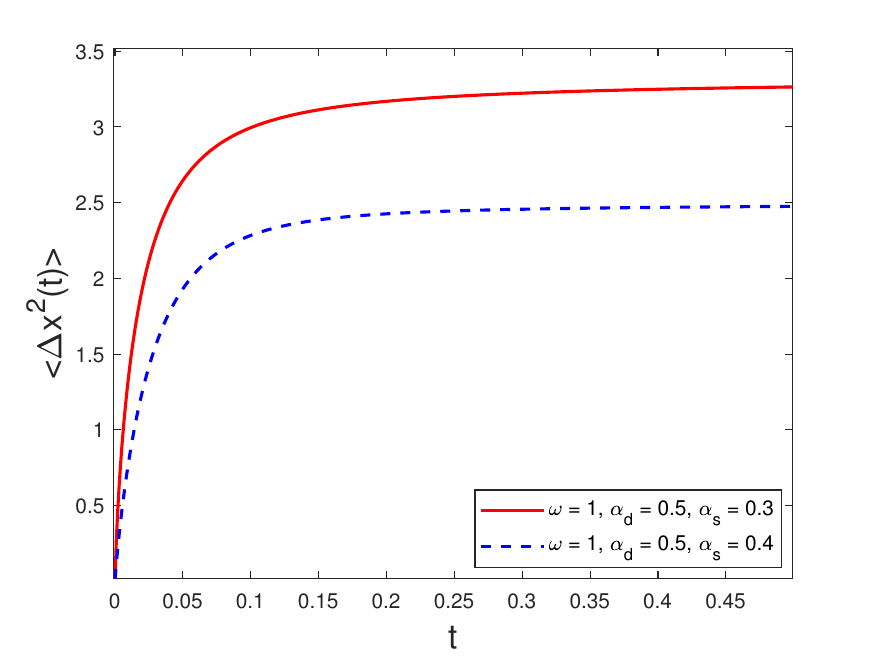} \\
(a) & (b) 
\end{tabular}
\caption{The stationary distribution (\ref{PL-PL-PDF-E1}) in panel (a) and the MSD (\ref{PL-PL-MSD-E1}) in panel (b) of the intermittent CTRW with the power-law jump and reset WTDs. The values of relevant model parameters are provided in the legend.}
\label{Fig-PL-PL-E}
\end{figure}

An interesting phenomenon is that the stationary distribution (\ref{PL-PL-PDF-E1}) for the power-law jump and reset WTDs in Fig. \ref{Fig-PL-PL-E}(a) is very similar to that for the exponential jump and reset WTDs, while the MSD (\ref{PL-PL-MSD-E1}) exhibits a Mittag-Leffer type evolution, reaching equilibrium within an extremely short time and decreasing with increasing the reset power-law distribution exponent $\alpha_s$ in Fig. \ref{Fig-PL-PL-E}(b).

\subsection{Gaussian distributed jump length}
\label{Gaussian distributed}
The Gaussian distribution is one of the most important and frequently used probability distributions. In the CTRW framework, it is often used as the distribution for jump lengths in random walks when studying phenomena of subdiffusion (and normal diffusion).
Here we assume the PDF of jump is Gaussian distributed as follows:
\begin{flalign}
\begin{split}
\Lambda(\xi)
=\frac{1}{\sqrt{2\pi\sigma^2}}\exp\{-\frac{\xi^2}{2\sigma^2}\},
\label{Jump-length-G}
\end{split}
\end{flalign}
and its Fourier transform can be approximated in the form of power series (up to $k^2$-order infinitesimal)
\cite{{MKS1998}}
\begin{flalign}
\begin{split}
\Lambda(k)
=\exp\{-\frac{\sigma^2 k^2}{2}\}\simeq 1-\frac{\sigma^2 k^2}{2},
\label{lambda-eq1}
\end{split}
\end{flalign}
for $\sigma k\ll 1$. 
Since the exact Fourier transform of the Gaussian distribution has the form of an exponential function, this poses a certain challenge for the investigation of particle distributions over the entire spatial domain within the CTRW framework. In order to investigate the diffusion behavior of particles, the consideration of longer jumps is a frequently used approximation method for such investigations. That is, the condition $\sigma k\ll 1$ implies $\sigma\ll |\xi|$.

\begin{figure}[htbp]
\centering
\includegraphics[width=0.8\linewidth]{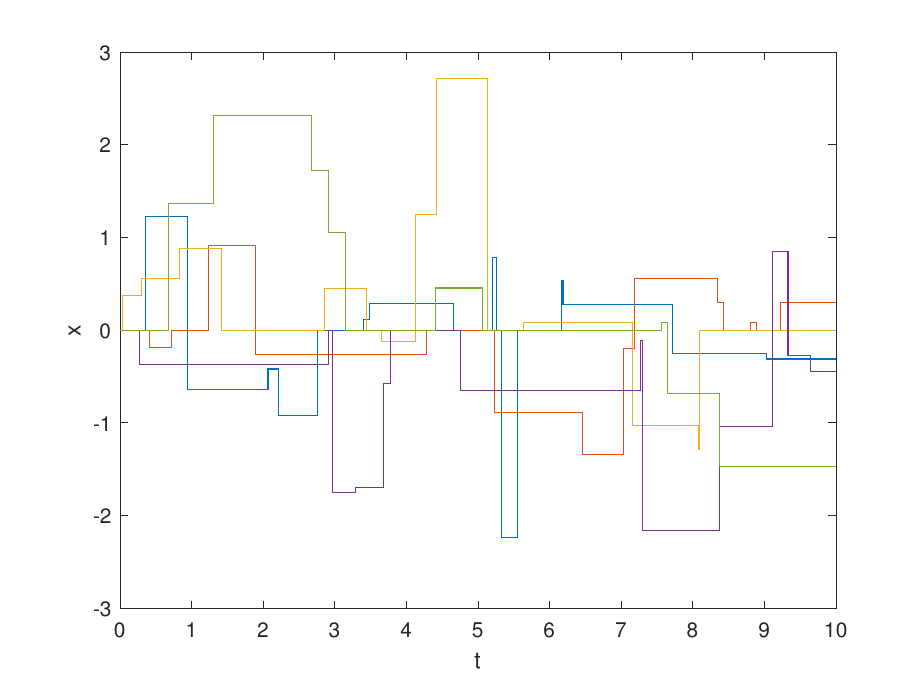} 
\caption{The particle trajectories of intermittent CTRW under stochastic resetting to original position.
 The exponential jump and reset WTDs with frequency parameters $\lambda_d=1.2$ and $\lambda_s=0.4$ individually and the Gaussian distributed jump length with standard deviation $\sigma=1$.}
\label{fig_2}
\end{figure}

Taking into account the most commonly used Gaussian distribution as the jump length, we have created a schematic diagram of the particle trajectory of intermittent CTRW under stochastic resetting (see Fig.~\ref{fig_2}) to illustrate the competition between jumps and resettings using the sample trajectory. From these sample trajectories, one can find that particles have a low probability of resetting to the origin and a high probability of maintaining their diffusion motion.

For the Gaussian distributed jump length the Eq. (\ref{master-eq0}) is written then as
\begin{flalign}
\begin{split}
u\rho(k,u)-\rho_0(k)\simeq &-\frac{\sigma^2}{2} M_d(u)k^2\rho(k,u)
+M_s(u)[\frac{1}{u}-\rho(k,u)].
\label{}
\end{split}
\end{flalign}
Taking inverse Fourier and Laplace transforms one then can obtain the governing equation with the Gaussian distributed jump length for arbitrary jump and reset WTDs
\begin{flalign}
\begin{split}
\frac{\partial\rho(x,t)}{\partial t}=\int_0^t M_s(t-\tau)[\delta(x)-\rho(x,\tau)] d\tau
 +\frac{\sigma^2}{2}\int_0^t M_d(t-\tau)\frac{\partial^2 \rho(x,\tau)}{\partial x^2} d\tau.
\label{Gaussian-ME-eq0}
\end{split}
\end{flalign}
 
Inserting Eq. (\ref{lambda-eq1}) into Eq. (\ref{ML-eq11}) with the initial condition $\rho_0(x)=\delta(x)$ and taking inverse Fourier transform, one can get the approximated tail of PDF of particles in Laplace space
\begin{flalign}
\begin{split}
\rho(x,u)\simeq\frac{1}{2u}\sqrt{\frac{2}{\sigma^2 }(\varphi^{-1}_d(u)-1)}\exp\left\{-\sqrt{\frac{2}{\sigma^2 }(\varphi^{-1}_d(u)-1)}\bigg|x\bigg|\right\},
\label{Gaussian-tail-PDF-eq0}
\end{split}
\end{flalign}
for $\sigma\ll|x|$.

 By solving for the first and second order derivatives of Eq. (\ref{ML-eq11}) with respect to the variable $k$ for the Gaussian distributed jump length one can obtain the first and second moments in Laplace space in the following
 relations
\begin{flalign}
\begin{split}
&\langle x(u)\rangle =i\frac{\partial}{\partial k}\rho(k,u)|_{k=0}= 0,\\
&\langle x^2(u)\rangle =i^2\frac{\partial^2}{\partial k^2}\rho(k,u)|_{k=0}= \frac{\sigma^2\varphi_d(u)}{u(1-\varphi_d(u))}.
\label{Gaussian-Second-M0}
\end{split}
\end{flalign}
That means the first moment $\langle x(t)\rangle =0$ for any distributions of waiting time to jump and reset in this case. Comparing the forms of the first two moments (\ref{E-MSD-eq0} ) in Laplace space for the exponential distributed jump length, it can be seen that the first two moments (\ref{Gaussian-Second-M0} ) in Laplace space in the Gaussian distribution case bears a strong resemblance to it. This is because both the Gaussian distribution and the exponential distribution have the similar first two moments, but higher-order moments will exhibit distinct behavior.

By inserting the Gaussian-distributed jump length (\ref{lambda-eq1}) into the propagation (\ref{ML-eq0}) of the intermittent CTRW, the distribution of FAT in Laplace space can be derived from its general relationship (\ref{Eq-FAT2}) as follows:
\begin{flalign}
\begin{split}
\chi(u)=\frac{1-\varphi_d(u)}{u\Psi(u)\exp{(\sqrt{\frac{2(1-\varphi_d(u))}{\sigma^2\varphi_d(u)}}\big|x_0\big|)}+\varphi_s(u)},
\label{Eq-FAT-G0}
\end{split}
\end{flalign}
which will be used to solve MFAT issues.

When we consider again the exponential jump and reset WTDs, applying the memory kernels (\ref{exponential-MK-eq0}) from Eq. (\ref{Gaussian-ME-eq0})
we can get
\begin{flalign}
\begin{split}
\frac{\partial\rho(x,t)}{\partial t}=\frac{\sigma^2 \lambda_d}{2}\frac{\partial^2 \rho(x,t)}{\partial x^2}+\lambda_s[\delta(x)-\rho(x,t)],
\label{}
\end{split}
\end{flalign}
which is similar as the diffusion equation with resetting in Ref. \cite{EM2011}. 

Inserting Eq. (\ref{lambda-eq1}) into Eq. (\ref{exp-PDF-eq0}) and taking inverse Fourier transform we can obtain the approximated tail of temporal PDF of particles in real space
\begin{flalign}
\begin{split}
 \rho(x,t)
&\simeq \frac{\sqrt{2^{-1}\lambda_s\lambda_d^{-1}}}{\sigma}\exp{\left[-\sqrt{2\lambda_s\lambda_d^{-1}} \frac{|x|}{\sigma}\right]}+\frac{1}{\sqrt{2\pi\lambda_d\sigma^2t}}\exp{(-\frac{x^2}{2\lambda_d\sigma^2t}-\lambda_s t)}\\
&-\frac{\sqrt{\lambda_s\pi^{-1}t^{-1}}}{2\lambda_d\sigma^2}\exp{(-\lambda_s t)}\int_{-\infty}^{\infty}\exp \bigg(-\frac{(x-x')^2}{2\lambda_d\sigma^2t}-\sqrt{2\lambda_s\lambda_d^{-1}} \frac{|x'|}{\sigma}\bigg)dx',
\label{}
\end{split}
\end{flalign}
which has an approximated stationary state 
\begin{flalign}
\begin{split}
\lim_{t\rightarrow \infty}\rho(x, t)\simeq  \frac{\sqrt{2^{-1}\lambda_s\lambda_d^{-1}}}{\sigma}\exp{\left[-\sqrt{2\lambda_s\lambda_d^{-1}} \frac{|x|}{\sigma}\right]},
\label{exponential-SS-eq0}
\end{split}
\end{flalign}
for $\sigma\ll|x|$.  
The stationary distribution in the domain of origin can be obtained by the relation (\ref {stationary-eq1}) expressed as the sum of a quadratic function in the position variable and $\delta(x)$  function, which differs the exact stationary distribution (\ref{PDF1-E})
for the exponential distributed jump length (the sum of the first-order Taylor series of the variables $|x|$ and $\delta(x)$  function). The possible reason for this lies in the difference between the exponential and Gaussian distributed jump length in the domain of origin or in the discrepancy between the higher-order terms of their Fourier transform functions with respect to the variable $k$.

Combing the Eq. (\ref{exponential-eq1}), from the Eqs. (\ref{Gaussian-Second-M0}) we can get the MSD (the second moment with the initial condition of particle $x(0)=0$) with inverse Laplace transform
\begin{flalign}
\begin{split}
\langle x^2(t)\rangle = \frac{\lambda_d}{\lambda_s}\sigma^2(1-\exp{(-\lambda_s t)}).
\label{MSD-Gaus1}
\end{split}
\end{flalign}

For the exponential jump and reset WTDs, the distribution of FAT (\ref{Eq-FAT-G0}) in Laplace space becomes  
\begin{flalign}
\begin{split}
\chi(u)=\frac{\lambda_s+u}{u\exp{\left(\frac{\sqrt{2}|x_0|}{\sigma\sqrt{\lambda_d}}\sqrt{\lambda_s +u}\right)}+\lambda_s},
\label{}
\end{split}
\end{flalign}
which results in the MFAT as the relation
\begin{flalign}
\begin{split}
<T>=\frac{1}{\lambda_s}\left[\exp{\left(\frac{\sqrt{2\lambda_s}|x_0|}{\sigma\sqrt{\lambda_d}}\right)}-1\right].
\label{Eq-FAT-G1}
\end{split}
\end{flalign}

\begin{figure}[htbp]
\centering
\begin{tabular}{cc}
\includegraphics[width=0.48\textwidth]{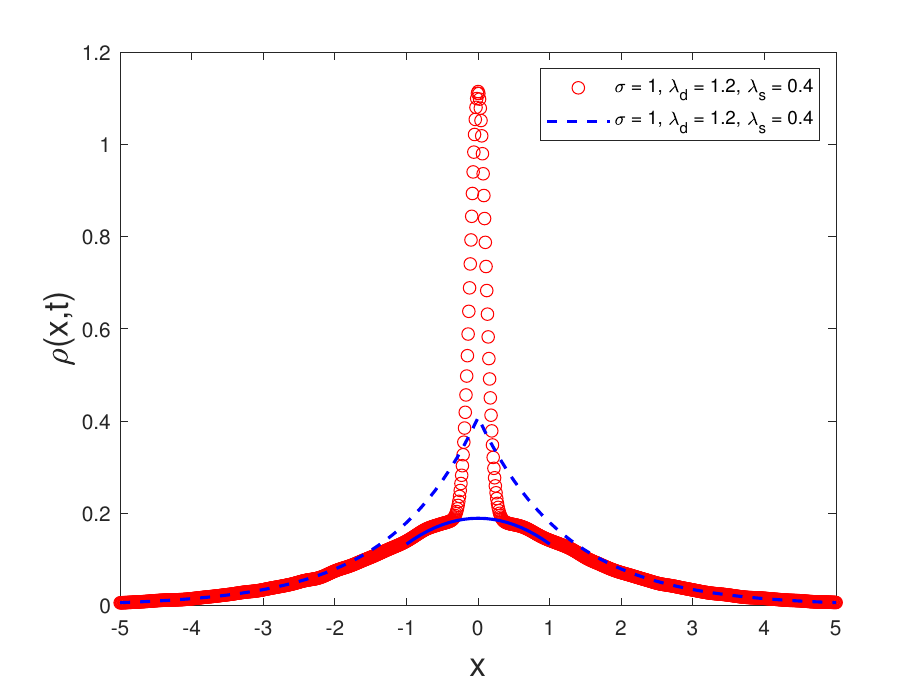} &
\includegraphics[width=0.48\textwidth]{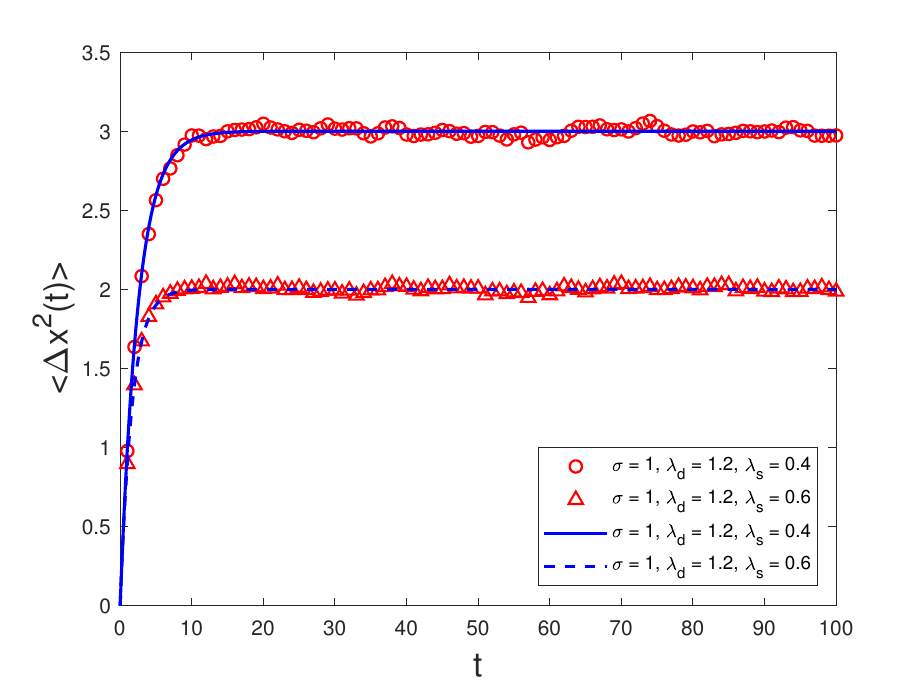} \\
(a) & (b) \\[0.3cm]
\multicolumn{2}{c}{\includegraphics[width=0.48\textwidth]{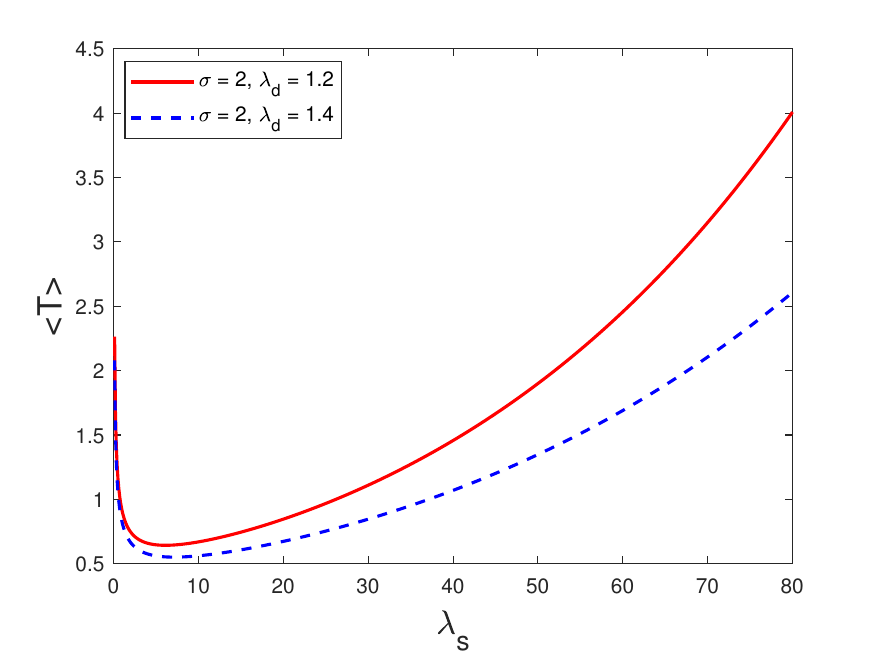}} \\
\multicolumn{2}{c}{(c)}
\end{tabular}
\caption{The same as in Figs. \ref{fig_1ab}, but for the Gaussian distributed jump length (\ref {Jump-length-G}). The full analytical results of the approximated stationary state with $\sigma\ll|x|$ (\ref {exponential-SS-eq0}) and close to $x = 0$ (\ref {stationary-eq1}) are shown with the results of simulations ($t=100$) in panel (a) as the solid curves. The theoretical results for the MSD (\ref {MSD-Gaus1}) are presented in panel (b) as the solid curves, together with the numerical results. The simulated sample size $N = 10^5$. In panel (c) the analytical results of MFAT (\ref{Eq-FAT-G1}) at the target position $x_0=1$ with the variation of the reset parameter $\lambda_s$.
The relevant parameters are given in the legend.}
	\label{fig_3}
\end{figure}

In Figs. \ref{fig_3}(a) and \ref{fig_3}(b) we perform simulations based on the jump lengths following Gaussian distribution to represent the PDF in stationary state and the MSD for intermittent CTRWs. The  analytical results of the approximated stationary state with $\sigma\ll|x|$ (\ref {exponential-SS-eq0}) and close to $x = 0$ (\ref {stationary-eq1}) agree with the result of corresponding simulation, and one can find the monotonicity and asymptotic stability of MSD (\ref{MSD-Gaus1}) from the simulation and analysis results.  Fig. \ref{fig_3}(c) shows how the MFAT (\ref{Eq-FAT-G1}) develops with the variation of the reset parameter $\lambda_s$, whereby MFAT reaches its minimum value at its stationary point.

Comparing the diffusion behavior of intermittent CTRW with the exponential distributed jump length reveals many similarities to Gaussian distribution, particularly in the MSD, even in the basic functional forms of stationary state distribution and MFAT. This is because the Fourier transform functions of exponential and Gaussian distributions have identical quadratic expansions in the variable $k$ and differ only in the higher-order terms. Consequently, the distribution properties of the diffusion systems become very similar far from the origin, while the distribution diverges near the origin. In MFAT, long-range jumping contributes significantly, so that the basic functional forms remain similar under both conditions. As can be seen from the following text, these properties exhibit the same behavior under other jump and reset WTDs conditions.

Considering the intermittent CTRW with the power-law jump and exponential reset WTDs, with the memory kernels (\ref{PL-E-MK-eq0}) and (\ref{PL-E-MK-eq1})
the governing equation (\ref{Gaussian-ME-eq0}) becomes
\begin{flalign}
\begin{split}
\frac{\partial\rho(x,t)}{\partial t}=&\frac{\sigma^2}{2 \tau_d} e^{-\lambda_s t}{}_{0}D_t^{1-\alpha} \left(e^{\lambda_s t}\frac{\partial^2 \rho(x,t)}{\partial x^2}\right)+\lambda_s[\delta(x)-\rho(x,t)].
\label{}
\end{split}
\end{flalign}
This expression agrees with the former result of Ref. \cite{KG2019} for the
subdiffusion with stochastic resetting.

Making use of the Eqs. (\ref{PL-E-eq0}) and (\ref{Gaussian-tail-PDF-eq0}) we obtain the asymptotic behavior of the PDF in the large time limit
\begin{flalign}
\begin{split}
\lim_{t\rightarrow \infty}\rho(x, t)\simeq\frac{1}{2}\sqrt{\frac{2}{\sigma^2 }\tau_d \lambda_s^{\alpha}}\exp\left\{-\sqrt{\frac{2}{\sigma^2 }\tau_d \lambda_s^{\alpha}}\bigg|x\bigg|\right\},
\label{Stationary-PDF-eq0}
\end{split}
\end{flalign}
for $\sigma\ll|x|$.

Inserting Eq. (\ref{PL-E-eq0}) into Eq. (\ref{Gaussian-Second-M0}), with inverse Laplace transform the MSD (the second moment) takes the form
\begin{flalign}
\begin{split}
\langle x^2(t)\rangle = \frac{\sigma^2}{\Gamma(\alpha)\tau_d \lambda_s^\alpha} \gamma(\alpha, \lambda_s t).
\label{Eq-MSD-G2}
\end{split}
\end{flalign}

By substituting the relationships (\ref{PL-E-eq0}) and (\ref{survival-eq3}) into expression (\ref{Eq-FAT-G0}), we obtain the distribution of FAT as
\begin{flalign}
\begin{split}
\chi(u)=\frac{u+\lambda_s}{u\exp{\left(\frac{\sqrt{2}|x_0|}{\sigma}\sqrt{\tau_d(u+\lambda_s)^\alpha }\right)}+\lambda_s},
\label{}
\end{split}
\end{flalign}
and the MFAT is given by
\begin{flalign}
\begin{split}
<T>=\frac{1}{\lambda_s}\left[\exp{\left(\frac{\sqrt{2\tau_d\lambda_s^\alpha}|x_0|}{\sigma}\right)}-1\right].
\label{Eq-FAT-G2}
\end{split}
\end{flalign}

\begin{figure}[htbp]
\centering
\begin{tabular}{cc}
\includegraphics[width=0.48\textwidth]{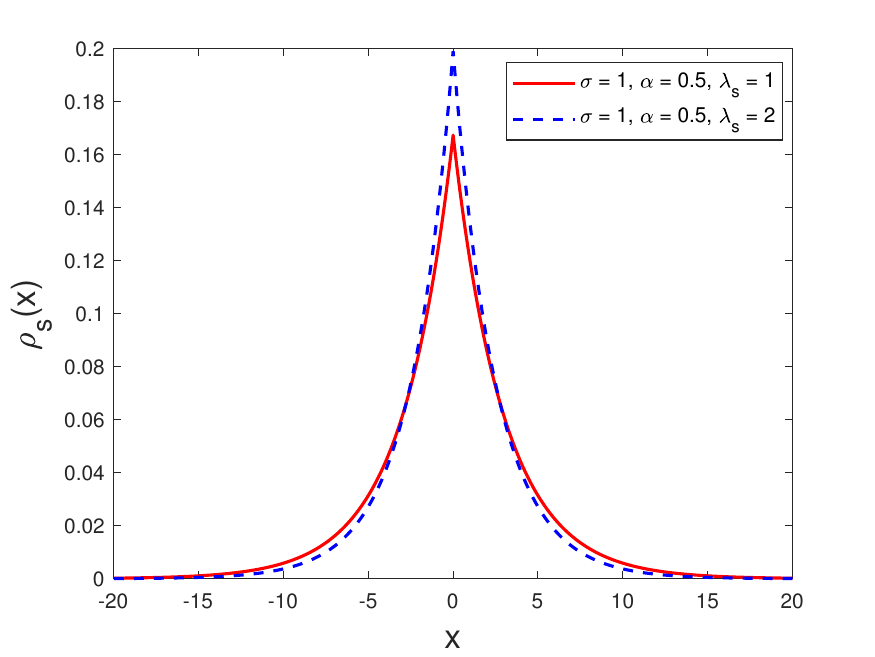} &
\includegraphics[width=0.48\textwidth]{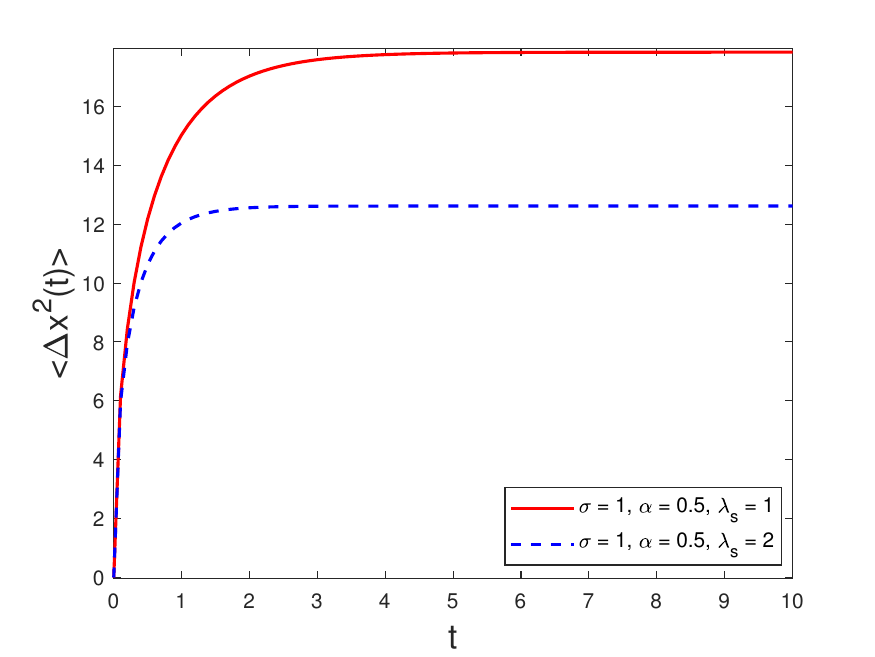} \\
(a) & (b) \\[0.3cm]
\multicolumn{2}{c}{\includegraphics[width=0.48\textwidth]{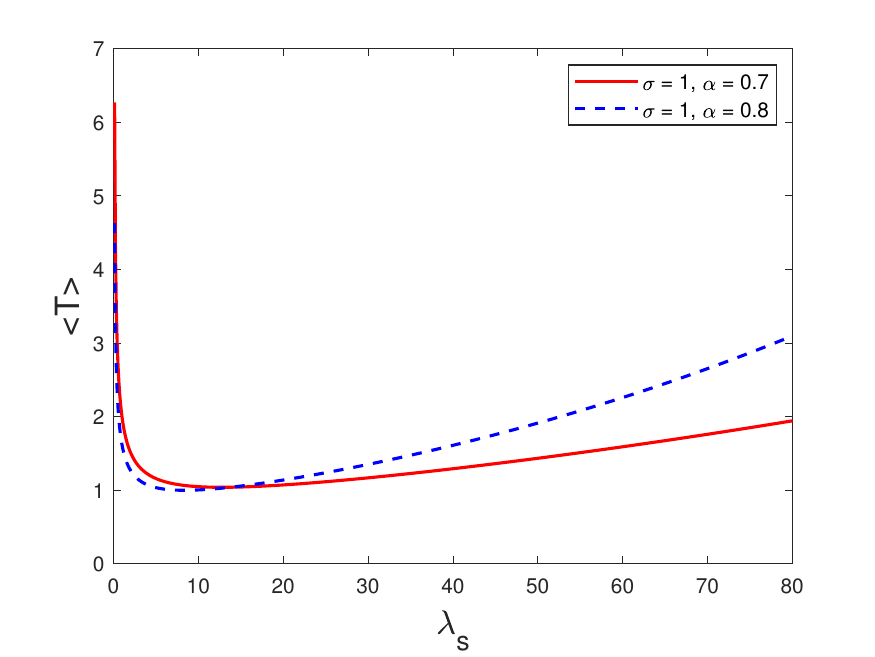}} \\
\multicolumn{2}{c}{(c)}
\end{tabular}
\caption{The same as in Figs. \ref{Fig-E2}, but for the Gaussian distributed jump length (\ref {Jump-length-G}). The analytical solutions for the approximated stationary state with $\sigma\ll|x|$ (\ref {Stationary-PDF-eq0}) in panel (a) and the MSDs (\ref {Eq-MSD-G2}) in panel (b) for the intermittent CTRWs with power-law jump WTD and exponential reset WTD. In panel (c) the analytical results of MFAT (\ref{Eq-FAT-G2}) at the target position $x_0=5$. The values of relevant model parameters are provided in the legend.}
\label{Fig-G2}
\end{figure}

Compared to the diffusion of intermittent CTRWs and the behavior of FAT under an exponential distributed jump length, the Gaussian distribution leads to similar properties in Figs. \ref{Fig-G2}(a), \ref{Fig-G2}(b) and \ref{Fig-G2}(c). For example, the second moment decreases as the reset parameter $\lambda_s$ increases. By adjusting this parameter, MFAT can be minimized to achieve optimal search efficiency.

In the case of exponential jump and power-law
reset WTDs, using the memory kernels (\ref{E-PL-MK-eq0}) and (\ref{E-PL-MK-eq1}) into the governing equation (\ref{Gaussian-ME-eq0}) leads to the result
\begin{flalign}
\begin{split}
\frac{\partial\rho(x,t)}{\partial t}=\frac{\sigma^2 \lambda_d}{2}\frac{\partial^2 \rho(x,t)}{\partial x^2}+\tau_s^{-1}e^{-\lambda_d t}{}_{0}D_t^{1-\alpha} \left(e^{\lambda_d t}[\delta(x)-\rho(x,t)]\right).
\label{}
\end{split}
\end{flalign}

Inserting Eq. (\ref{E-PL-eq0}) into Eq. (\ref{Gaussian-tail-PDF-eq0}) we can obtain an approximated stationary state using the previous method in Eq. (\ref{PDF1-E})
\begin{flalign}
\begin{split}
\lim_{t\rightarrow \infty}\rho(x, t)\simeq\frac{1}{2}\sqrt{\frac{2}{\sigma^2 }\tau_s^{-1} \lambda_d^{-\alpha}}\exp\left\{-\sqrt{\frac{2}{\sigma^2 }\tau_s^{-1} \lambda_d^{-\alpha}}\bigg|x\bigg|\right\},
\label{Eq-PDF-G3}
\end{split}
\end{flalign}
for $\sigma\ll|x|$.

Using Eq. (\ref{E-PL-eq0}) into the Eq. (\ref{Gaussian-Second-M0}) one get an asymptotic expansion in the
limit $u\rightarrow 0$ in the Laplace space
\begin{flalign}
\begin{split}
\langle x^2(u)\rangle
=\frac{\sigma^2\lambda_d \tau_s u^{-1}}{(u+\lambda_d)^{1-\alpha}+\tau_s u}\simeq \frac{\sigma^2\lambda_d \tau_s u^{-1}}{(u+\lambda_d)^{1-\alpha}},
\end{split}
\end{flalign}
with inverse Laplace transform one then can get the approximated MSD (the second moment) in the large time limit
\begin{flalign}
\begin{split}
\langle x^2(t)\rangle \simeq \frac{\sigma^2\lambda_d^{\alpha} \tau_s}{\Gamma(1-\alpha) } \gamma(1-\alpha, \lambda_d t),
\label{Eq-MSD-G3}
\end{split}
\end{flalign}
or in another form
\begin{flalign}
\begin{split}
\langle x^2(t)\rangle &\simeq \sigma^2\lambda_d \tau_s t^{1-\alpha}E_{1,2-\alpha,1-\alpha}(-\lambda_d t)\\
&\simeq \sigma^2\lambda_d^{\alpha} \tau_s .
\end{split}
\end{flalign}

 Applying the relationships (\ref{E-PL-eq0}) and (\ref{survival-eq3}), from Eq. (\ref{Eq-FAT-G0}) we get the distribution of FAT in Laplace space
\begin{flalign}
\begin{split}
\chi(u)=\frac{\tau_su+(u+\lambda_d)^{1-\alpha}}{u\tau_s\exp{\left(\frac{\sqrt{2}|x_0|}{\sigma\sqrt{\lambda_d\tau_s}}\sqrt{\tau_su+(u+\lambda_d)^{1-\alpha}}\right)}+(u+\lambda_d)^{1-\alpha}},
\label{}
\end{split}
\end{flalign}
and the MFAT at the target position $x_0$,
\begin{flalign}
\begin{split}
<T>=\frac{\tau_s}{\lambda_d^{1-\alpha}}\left[\exp{\left(\frac{\sqrt{2}|x_0|}{\sigma\sqrt{\tau_s\lambda_d^\alpha}}\right)}-1\right].
\label{Eq-MFAT-G3}
\end{split}
\end{flalign}

\begin{figure}[htbp]
\centering
\begin{tabular}{cc}
\includegraphics[width=0.48\textwidth]{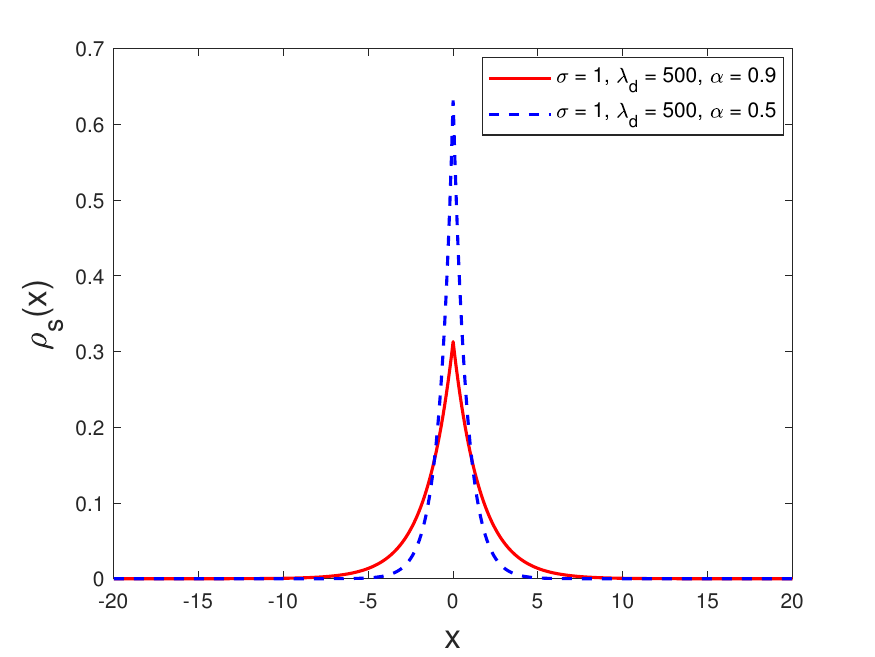} &
\includegraphics[width=0.48\textwidth]{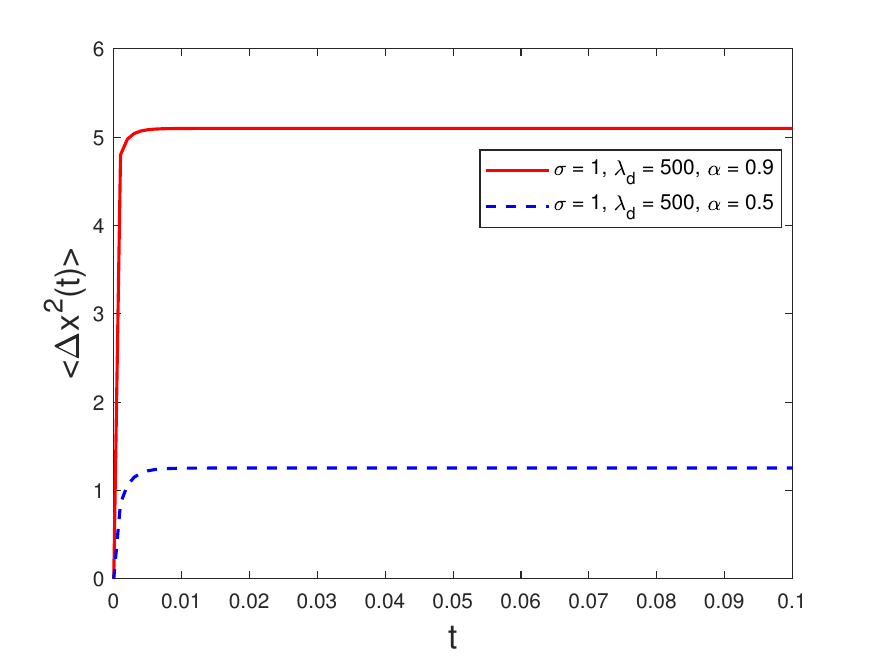} \\
(a) & (b) \\[0.3cm]
\multicolumn{2}{c}{\includegraphics[width=0.48\textwidth]{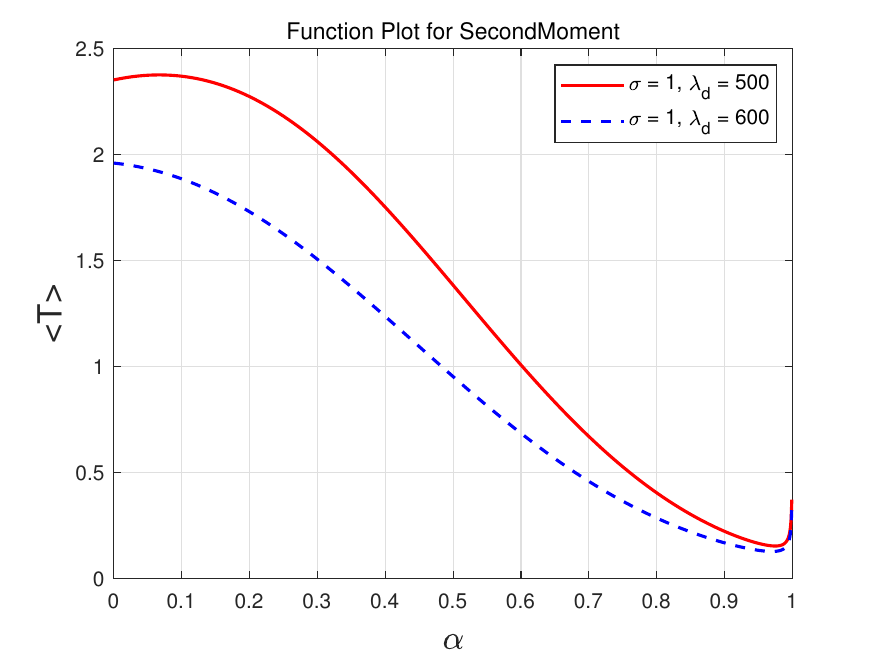}} \\
\multicolumn{2}{c}{(c)}
\end{tabular}
\caption{ The PDF of particles at stationary state with $\sigma\ll|x|$ (\ref {Eq-PDF-G3}) in panel (a) and the MSDs (\ref {Eq-MSD-G3}) in panel (b) for the intermittent CTRWs with power-law reset WTD. In panel (c) the analytical results of MFAT (\ref{Eq-MFAT-G3}) at the target position $x_0=5$. The values of relevant model parameters are provided in the legend.}
\label{Fig-G3}
\end{figure}

In Figs. \ref{Fig-G3}(a) and \ref{Fig-G3}(b) we have once again demonstrated the existence of on-equilibrium stationary state and the monotonically increasing property of MSD. At the same time, it can be observed that enhancing the parameter $\lambda_d$ of jump WTD also has a positive effect on the reduction of MFAT in Fig. \ref{Fig-G3}(c).

For the power-law jump and reset WTDs, substituting the memory kernels (\ref{PL-PL-MK-eq0}) into the governing equation (\ref{Gaussian-ME-eq0}) yields
\begin{flalign}
\begin{split}
\frac{\partial\rho(x,t)}{\partial t}\simeq&\frac{1}{\tau_{ds}(\alpha_d+\alpha_s)}{}_{0}D_t^{1-\alpha_d-\alpha_s}\bigg[\frac{\sigma^2 \alpha_d}{2}\frac{\partial^2 \rho(x,t)}{\partial x^2}+\alpha_s(\delta(x)-\rho(x,t))\bigg]
\label{Eq-GOV-G4}
\end{split}
\end{flalign}
for $0<\alpha_d+\alpha_s<1$.

By substituting equation (\ref{PL-PL-eq0}) into equation (\ref{Gaussian-tail-PDF-eq0}), we obtain an approximate steady state
\begin{flalign}
\begin{split}
\lim_{t\rightarrow \infty}\rho(x, t)\simeq  \frac{\sqrt{2^{-1}\alpha_s\alpha_d^{-1}}}{\sigma}\exp{\left[-\sqrt{2\alpha_s\alpha_d^{-1}} \frac{|x|}{\sigma}\right]},
\label{Eq-PDF-G4}
\end{split}
\end{flalign}
for $\sigma\ll|x|$.

The MSD can also be derived from equations (\ref{Gaussian-Second-M0}) by combining Eq.(\ref{PL-PL-eq0}) with the inverse Laplace transform
\begin{flalign}
\begin{split}
\langle x^2(t)\rangle \simeq \frac{\sigma^2\alpha_d}{\tau_{ds}(\alpha_d+\alpha_s)} t^{\alpha_d+\alpha_s}E_{\alpha_d+\alpha_s,\alpha_d+\alpha_s+1}\left(-\frac{\alpha_st^{\alpha_d+\alpha_s}}{\tau_{ds}(\alpha_d+\alpha_s)} \right).
\label{Eq-MSD-G4}
\end{split}
\end{flalign}

 Applying the relations (\ref{PL-PL-eq0}) and (\ref{survival-eq3}), we obtain the distribution of FAT in Laplace space from Eq. (\ref{Eq-FAT-G0}),
\begin{flalign}
\begin{split}
\chi(u)=\frac{(\alpha_d+\alpha_s)\tau_{ds}u^{\alpha_d+\alpha_s}+\alpha_s}{(\alpha_d+\alpha_s)\tau_{ds}u^{\alpha_d+\alpha_s}\exp{\left(\frac{\sqrt{2}|x_0|}{\sigma\sqrt{\alpha_d}}\sqrt{(\alpha_d+\alpha_s)\tau_{ds}u^{\alpha_d+\alpha_s}+\alpha_s}\right)}+\alpha_s}.
\label{}
\end{split}
\end{flalign}
As was to be expected, the MFAT is divergent in this situation,
\begin{flalign}
\begin{split}
<T>=\infty.
\label{}
\end{split}
\end{flalign}

\begin{figure}[htbp]
\centering
\begin{tabular}{cc}
\includegraphics[width=0.48\textwidth]{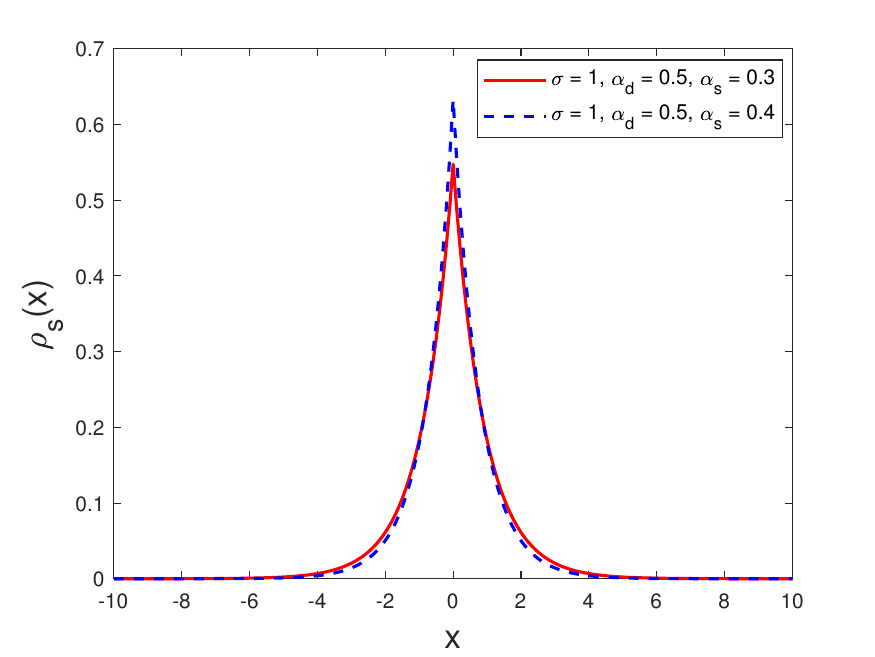} &
\includegraphics[width=0.48\textwidth]{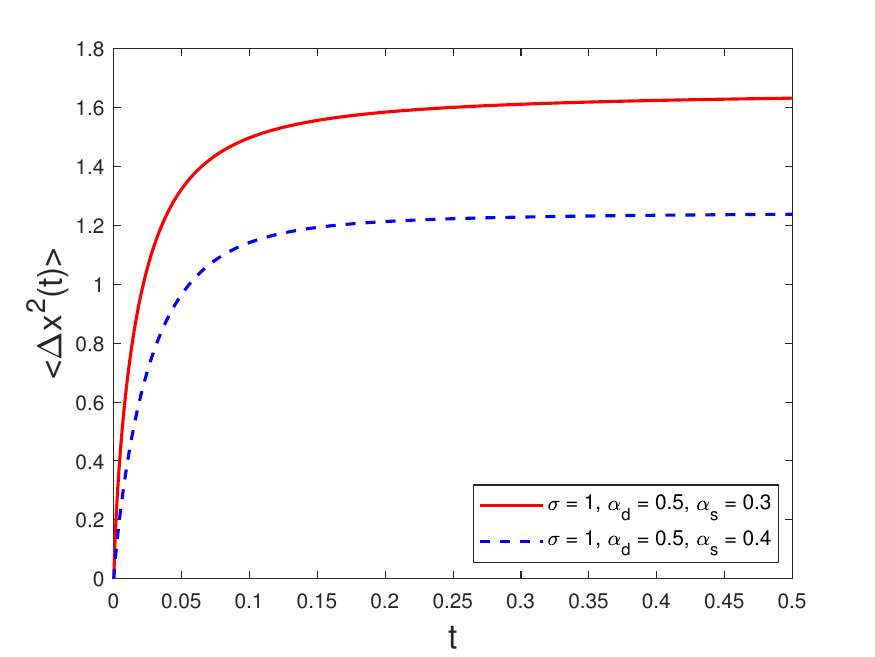} \\
(a) & (b) 
\end{tabular}
\caption{  The shape of the stationary state  with $\sigma\ll|x|$ (\ref {Eq-PDF-G4}) in panel (a) and the MSDs (\ref {Eq-MSD-G4}) in panel (b) for the intermittent CTRWs with power-law jump and reset WTDs. The values of relevant model parameters are provided in the legend.}
\label{Fig-G4}
\end{figure}

Similarity of the stationary distribution in Eq. (\ref{Eq-PDF-G4}) for the power-law jump and reset WTDs in Fig. \ref{Fig-G4}(a) reappears with that in Eq. (\ref{exponential-SS-eq0}) for the exponential jump and reset WTDs, the MSD decreases with increasing the reset power-law distribution exponent $\alpha_s$ in Fig. \ref{Fig-G4}(b), and diffusion dispersion is limited by the resetting.

\section{Analysis, comparison, and simulation}
\label{Comparison}

The fundamental motivation for intermittent CTRWs under the renewal reset mechanism is to treat both the diffusion motion of the particle and the random reset as system renewal. The shorter waiting time between the jump and reset waiting times for the particle is considered the system renewal time. Therefore, after each system renewal—regardless of whether the particle jumps or resets—the jump and reset waiting times are restarted (resampled in the simulation). The restarted jump and reset waiting times are then compared again to determine their next renewal action. 
This differs from the two reset mechanisms (complete and incomplete resettings) based on the CTRWs described in Refs. \cite{KG2019, BS2020}. The first model of complete resetting  should reset the diffusion motion of the particle, including its position and waiting time. (To our understanding) diffusion does not seem to have any influence on the reset. The second model, the incomplete resetting, resets the position of the diffusion particle while its jump waiting time remain unchanged. The particle diffusion also has no influence on the resetting.

The most similar effect that these different reset mechanisms above have on CTRW is the presence of non-equilibrium stationary states, which is one of the fundamental characteristics of stochastic resetting. Both the renewal resetting (in this article) and the complete resetting exhibit finite MFAT for the exponential reset WTD, which provides a reliable theoretical basis for search efficiency in practical search application scenarios. In contrast, the MFAT diverges for incomplete resetting under such conditions. The advantage of the renewal resetting lies not only in its finite MFAT for the exponential reset WTD, but also in its finite MFAT in the intermittent CTRW with exponential jump WTD even for power-law reset WTD. However, the MFAT diverges for power-law jump and reset WTDs.

It should be noted that in Ref. \cite{NG2016}, for diffusion with power-law resetting in the range of the power-law distribution exponent $0<\alpha<1$, i.e., for non-Markovian resetting with memory, neither a stationary state nor a finite MFAT exists, which differs from renewal reset in this article. The main reason for this is that the mean waiting time between successive resetting events diverges and particle diffusion has no influence on the resetting in Ref. \cite{NG2016}. 
In the renewal reset model, however, the diffusion movement of the particles can "reduce" the memory effect of the power-law distribution of reset waiting time. This is the distinguishing feature from the reset models in other publications,
such as complete and incomplete resettings with power-law distribution \cite{SS2022}, and random walks under resetting with constant probability at each renewal \cite{MC2016}.

Through our analysis, we find that under Markovian resetting (i.e., exponential  distributed reset time), the renewal reset mechanism can be reduced to the first model of complete resetting in Ref. \cite{KG2019, MMSC2021}. This could be due to the memoryless property of the exponential distribution. As for the power-law jump and reset WTDs leaded to the existence of non-equilibrium stationary states in the intermittent CTRWs with the renewal resetting, the probable cause is that after each system renewal, both the particle jump and reset waiting times are restarted.

 Under various conditions where the jump length follows to an exponential and Gaussian distribution, the intermittent CTRWs exhibits similar diffusion behavior and MFAT properties. This is due to the high consistency between exponential and Gaussian distributions in long-range regions. For the exponential distribution, we derive exact analytical expressions for the  stationary distribution, MSD, and MFAT (in Laplace space), thereby providing significant theoretical value for the study of renewal reset processes. For the Gaussian distribution—the most critical distribution in practical applications—the approximate analytical results presented here for the stationary distribution, MSD, and MFAT are of considerable practical importance.

To validate the accuracy and rationality of the theoretical results for the renewal reset mechanism, the article presents simulations of intermittent CTRWs with jump lengths following exponential and Gaussian distributions for the exponential jump and reset WTDs in Figs. \ref{fig_1ab} and \ref{fig_3}. The specific programming steps are as follows:

1. Initialize all particles at the origin and at time zero. For each particle, take a random sample separately for the jump waiting time and the reset waiting time. For the exponential jump and reset WTDs, sample according to the standard exponential distribution with different parameters.

2. Compare the jump waiting time and the reset waiting time of the random particle. If the jump waiting time is less than the reset waiting time, randomly sample a jump length and add it to the particle's position variable. For jump length that follows an exponential distribution, select randomly from the standard exponential distribution, and select a probability of $1/2$ to determine the jump direction of the particle to the left or right (jump length sample multiplied by $\pm 1$). If the jump length follows a Gaussian distribution, select a value from the Gaussian distribution. If the jump waiting time exceeds the reset waiting time, reset the position of the particle to zero. Add the minimum waiting time to the system time.

3. Resample the jump waiting time and reset waiting time for the particle whose positions have been updated, then repeat step 2.

4. If the system time exceeds the total simulation time, terminate the process.

\section{Conclusions} \label{sec-disc}

Stochastic resetting as a practical and efficient search strategy can be applied in searching processes of all kinds in complex environments, such as animal foraging, protein identification in DNA, and data mining.

In general, the diffusion propagation of particles under random resetting repeatedly returns to its initial state, which is highly likely to result in non-equilibrium stationary state and finite MFAT. This is advantageous for researcher in improving search efficiency.

Some publications are devoted to exploring a general approach to stochastic resetting, while numerous reset mechanisms with different underlying processes have been established, particularly for CTRWs, which are made more interesting by the question of whether the waiting time is restarted.

Based on the competition between jumping and resetting, this article proposes and investigates intermittent CTRWs under renewal resetting, using the smaller waiting time for jump and reset as the renewal time, where the non-equilibrium stationary states exist for exponential and power-law WTDs for both jump and reset.

Through the discussion of Markovian resetting for the intermittent CTRWs, we find that it is consistent with the first model of subdiffusion with stochastic resetting in Ref. \cite{KG2019}. By introducing the stationary reset probability, the result of stationary state in intermittent CTRWs under renewal reset mechanism weakly similar to that previously obtained in Ref. \cite{MC2016}, but in fact the reset probability in our model is time-dependent with memory. 

For the exponential and Gaussian distributed jump lengths, we derive the analytical expressions for the stationary distributions, MSDs, and MFATs, thereby providing significant theoretical and practical values for the renewal reset processes. These theoretical results can also find practical applications in many disciplines such as foraging and search strategies \cite{ref75,ref76} and robotic search \cite{ref78,ref79}.

Compared to existing reset models in various references, our intermittent CTRW model based on the competition between jumping and resetting offers a wide range of theoretical innovations:

1. Our intermittent CTRW model under renewal reset mechanism can be applied to any distrbutions of jump and reset times, including but not limited to those with exponential and power-law distributions. This highlights the versatility and universality of our reset model for various search scenarios.

2. For exponential reset times, many existing Markovian reset models can be regarded as special cases of the intermittent CTRW under renewal reset mechanism, which clearly demonstrates the reliability and correctness of our reset model.

3. This paper shows that the intermittent CTRW model with the renewal competition can improve search efficiency. The exponential jump and power-law reset WTDs exhibit a finite MFAT, this conclusion stands in stark contrast to the results in Ref. \cite{NG2016}, where neither a stationary state nor a finite MFAT exists.

4. In this work, the macroscopic fractional governing equations with renewal-resetting from the framework of intermittent CTRWs are derived using rigorous mathematical methods for both power-law jump times and reset times. 
The corresponding fractional dynamic systems in Eqs. (\ref{PL-PL-GOV-E1}) and (\ref{Eq-GOV-G4}) with non-equilibrium stationary states, as well as their MSDs in Eqs. (\ref{PL-PL-MSD-E1}) and (\ref{Eq-MSD-G4}), exhibit distinct fractal properties rather than merely reproducing the results of existing reset models.

\roles{}
\textbf{Guohua Li}: Conceptualization, Methodology, Writing – original draft, Editing, Visualization,  Validation, Investigation. \textbf{Hong Zhang}: Conceptualization, Methodology, Reviewing, Validation, Investigation, Formal analysis, Funding acquisition. \textbf{Yuexiong Liu}: Reviewing, Validation, Formal analysis

\data{No data was used for the research described in the article.}

\ack{The authors wish to thank the anonymous reviewers for
their valuable suggestions leading to the improvement of
the paper.}

\begin{appendices}

\section{Approaches and definitions of CTRWs} \label{sec-ctrw-overview}

 Here we assume that the jumps $\{\xi_1, \xi_2, \dots\}$ are independent identically distributed (iid) random variables and the jumps occur at random times $\{T_1, T_2, \dots \}$, so that waiting time intervals between the jumps $\theta_n = T_n-T_{n-1}$ can also be assumed as iid random variables. Denoting the position of a particle at time $t$ by $X(t)$, and with the initial position being $X(0) = 0$ one has 
\begin{flalign}
\begin{split}
X(t)=\sum_{i=1}^{N(t)}\xi_{i},
\label{ctrw-par}
\end{split}
\end{flalign}
where the renewal process $N(t)$ is the number of jumps up to time $t$.  It can be defined in terms of the random time $T_n$ as   
\begin{flalign}
\begin{split}
N(t)=\max\left\{ n\geq{0}:T_n\leq{t} \right\}.
\label{ctrw-jum-num}
\end{split}
\end{flalign}
The sequence of particle positions $X(t)$ is called a CTRW.

If the waiting-time intervals $\theta_n$ and jumps $\xi_n$ to be dependent, let $\phi(\xi,t)$ denote their joint PDF, then the jump-PDF is
\begin{flalign}
\begin{split}
\Lambda(\xi)=\int_0^{\infty}\phi(\xi,t)dt,\label{ctrw-jum}
\end{split}
\end{flalign}
and the PDF of the WTD is
\begin{flalign}
\begin{split}
\psi(t)=\int_{-\infty}^{\infty}\phi(\xi,t)d\xi.
\label{ctrw-wait}
\end{split}
\end{flalign}
 Let $\Psi(t)$ be the probability that no jump has occurred up to time $t$, that is, the survival probability
\begin{flalign}
\begin{split}
\Psi(t)=1-\int_0^t\psi(\tau)d\tau.
\label{ctrw-surv}
\end{split}
\end{flalign}

Let $\rho(x,t)$ be the PDF of the particle located at position $x$ at time $t$. The balance equations  can be given as \cite{MFH2010}
\begin{flalign}
\begin{split}
\rho(x,t)=\rho_0(x)\Psi(t)+\int_0^t I(x,t-\tau)\Psi(\tau) d\tau,
\label{ctrw-evol-eq1}
\end{split}
\end{flalign}
and here $I(x,t)$ be the flux of particles that reach the position $x$ exactly at time $t$,
\begin{flalign}
\begin{split}
I(x,t)=\int_{-\infty}^{\infty}\rho_0(x-\xi)\phi(t,\xi)d\xi +\int_0^t d\tau \int_{-\infty}^{\infty} I(x-\xi,t-\tau)\phi(\tau, \xi) d\xi.
\label{ctrw-evol-eq2}
\end{split}
\end{flalign}
where $\rho_0(x)$ is the initial PDF. 

In Fourier-Laplace space one then obtains the coupled Montroll-Weiss equation \cite{KS2011, MFH2010}
\begin{flalign}
\begin{split}
\rho(k,u)=\frac{\rho_0(k)[1-\psi(u)]}{u[1-\phi(k,u)]},
\label{ctrw-MW-eq}
\end{split}
\end{flalign}
here $\rho_0(k)$ is the Fourier transform of $\rho_0(x)$ given by\cite{JD2008} 
\begin{flalign}
\begin{split}
 \rho_0(k)=\int_{-\infty }^{\infty} e^{-ikx}\rho_0(x)dx,
\end{split}
\end{flalign}
  $\psi(u)$ is the Laplace transform of $\psi(t)$ given by
 \begin{flalign}
\begin{split}
 \psi(u) = \mathcal{L} (\psi(t)) = \int _{0} ^{\infty} e^{-st}\psi(t)dt, 
\end{split}
\end{flalign}
and $\rho(k,u)$ is the Fourier-Laplace transform of $\rho (x, t)$ defined as 
\begin{flalign}
\begin{split}
 \rho(k,u) =\int _{0} ^{ \infty }dt\int_{-\infty}^{\infty} e^{-st- ikx} \rho(x,t)dx. 
\end{split}
\end{flalign} 
The evolution equation of the PDF then is written  as  \cite{ MFH2010}
\begin{flalign}
\begin{split}
\rho(x,t)=\rho_0(x)\Psi(t)+\int_0^t\int_{-\infty}^{\infty}\rho(x-\xi,\tau)\phi(\xi,t-\tau)d\xi d\tau.
\label{ctrw-evol-eq}
\end{split}
\end{flalign}

\end{appendices}

\bibliographystyle{iopart-num} 
\bibliography{reference}

\end{document}